# Transverse momentum cross section of $e^+e^-$ pairs in the $Z$-boson region from $p\bar{p}$ collisions at $\sqrt{s} = 1.96$ TeV

T. Aaltonen,[21] B. Álvarez González[z],[9] S. Amerio,[40] D. Amidei,[32] A. Anastassov[x],[15] A. Annovi,[17] J. Antos,[12] G. Apollinari,[15] J.A. Appel,[15] T. Arisawa,[54] A. Artikov,[13] J. Asaadi,[49] W. Ashmanskas,[15] B. Auerbach,[57] A. Aurisano,[49] F. Azfar,[39] W. Badgett,[15] T. Bae,[25] A. Barbaro-Galtieri,[26] V.E. Barnes,[44] B.A. Barnett,[23] P. Barria[hh],[42] P. Bartos,[12] M. Bauce[ff],[40] F. Bedeschi,[42] S. Behari,[23] G. Bellettini[gg],[42] J. Bellinger,[56] D. Benjamin,[14] A. Beretvas,[15] A. Bhatti,[46] D. Bisello[ff],[40] I. Bizjak,[28] K.R. Bland,[5] B. Blumenfeld,[23] A. Bocci,[14] A. Bodek,[45] D. Bortoletto,[44] J. Boudreau,[43] A. Boveia,[11] L. Brigliadori[ee],[6] C. Bromberg,[33] E. Brucken,[21] J. Budagov,[13] H.S. Budd,[45] K. Burkett,[15] G. Busetto[ff],[40] P. Bussey,[19] A. Buzatu,[31] A. Calamba,[10] C. Calancha,[29] S. Camarda,[4] M. Campanelli,[28] M. Campbell,[32] F. Canelli,[11,15] B. Carls,[22] D. Carlsmith,[56] R. Carosi,[42] S. Carrillo[m],[16] S. Carron,[15] B. Casal[k],[9] M. Casarsa,[50] A. Castro[ee],[6] P. Catastini,[20] D. Cauz,[50] V. Cavaliere,[22] M. Cavalli-Sforza,[4] A. Cerri[f],[26] L. Cerrito[s],[28] Y.C. Chen,[1] M. Chertok,[7] G. Chiarelli,[42] G. Chlachidze,[15] F. Chlebana,[15] K. Cho,[25] D. Chokheli,[13] W.H. Chung,[56] Y.S. Chung,[45] M.A. Ciocci[hh],[42] A. Clark,[18] C. Clarke,[55] G. Compostella[ff],[40] M.E. Convery,[15] J. Conway,[7] M.Corbo,[15] M. Cordelli,[17] C.A. Cox,[7] D.J. Cox,[7] F. Crescioli[gg],[42] J. Cuevas[z],[9] R. Culbertson,[15] D. Dagenhart,[15] N. d'Ascenzo[w],[15] M. Datta,[15] P. de Barbaro,[45] M. Dell'Orso[gg],[42] L. Demortier,[46] M. Deninno,[6] F. Devoto,[21] M. d'Errico[ff],[40] A. Di Canto[gg],[42] B. Di Ruzza,[15] J.R. Dittmann,[5] M. D'Onofrio,[27] S. Donati[gg],[42] P. Dong,[15] M. Dorigo,[50] T. Dorigo,[40] K. Ebina,[54] A. Elagin,[49] A. Eppig,[32] R. Erbacher,[7] S. Errede,[22] N. Ershaidat[dd],[15] R. Eusebi,[49] S. Farrington,[39] M. Feindt,[24] J.P. Fernandez,[29] R. Field,[16] G. Flanagan[u],[15] R. Forrest,[7] M.J. Frank,[5] M. Franklin,[20] J.C. Freeman,[15] Y. Funakoshi,[54] I. Furic,[16] M. Gallinaro,[46] J.E. Garcia,[18] A.F. Garfinkel,[44] P. Garosi[hh],[42] H. Gerberich,[22] E. Gerchtein,[15] S. Giagu,[47] V. Giakoumopoulou,[3] P. Giannetti,[42] K. Gibson,[43] C.M. Ginsburg,[15] N. Giokaris,[3] P. Giromini,[17] G. Giurgiu,[23] V. Glagolev,[13] D. Glenzinski,[15] M. Gold,[35] D. Goldin,[49] N. Goldschmidt,[16] A. Golossanov,[15] G. Gomez,[9] G. Gomez-Ceballos,[30] M. Goncharov,[30] O. González,[29] I. Gorelov,[35] A.T. Goshaw,[14] K. Goulianos,[46] S. Grinstein,[4] C. Grosso-Pilcher,[11] R.C. Group[53],[15] J. Guimaraes da Costa,[20] S.R. Hahn,[15] E. Halkiadakis,[48] A. Hamaguchi,[38] J.Y. Han,[45] F. Happacher,[17] K. Hara,[51] D. Hare,[48] M. Hare,[52] R.F. Harr,[55] K. Hatakeyama,[5] C. Hays,[39] M. Heck,[24] J. Heinrich,[41] M. Herndon,[56] S. Hewamanage,[5] A. Hocker,[15] W. Hopkins[g],[15] D. Horn,[24] S. Hou,[1] R.E. Hughes,[36] M. Hurwitz,[11] U. Husemann,[57] N. Hussain,[31] M. Hussein,[33] J. Huston,[33] G. Introzzi,[42] M. Iori[jj],[47] A. Ivanov[p],[7] E. James,[15] D. Jang,[10] B. Jayatilaka,[14] E.J. Jeon,[25] S. Jindariani,[15] M. Jones,[44] K.K. Joo,[25] S.Y. Jun,[10] T.R. Junk,[15] T. Kamon[25],[49] P.E. Karchin,[55] A. Kasmi,[5] Y. Kato[o],[38] W. Ketchum,[11] J. Keung,[41] V. Khotilovich,[49] B. Kilminster,[15] D.H. Kim,[25] H.S. Kim,[25] J.E. Kim,[25] M.J. Kim,[17] S.B. Kim,[25] S.H. Kim,[51] Y.K. Kim,[11] Y.J. Kim,[25] N. Kimura,[54] M. Kirby,[15] S. Klimenko,[16] K. Knoepfel,[15] K. Kondo,[54,*] D.J. Kong,[25] J. Konigsberg,[16] A.V. Kotwal,[14] M. Kreps,[24] J. Kroll,[41] D. Krop,[11] M. Kruse,[14] V. Krutelyov[c],[49] T. Kuhr,[24] M. Kurata,[51] S. Kwang,[11] A.T. Laasanen,[44] S. Lami,[42] S. Lammel,[15] M. Lancaster,[28] R.L. Lander,[7] K. Lannon[y],[36] A. Lath,[48] G. Latino[hh],[42] T. LeCompte,[2] E. Lee,[49] H.S. Lee[q],[11] J.S. Lee,[25] S.W. Lee[bb],[49] S. Leo[gg],[42] S. Leone,[42] J.D. Lewis,[15] A. Limosani[t],[14] C.-J. Lin,[26] M. Lindgren,[15] E. Lipeles,[41] A. Lister,[18] D.O. Litvintsev,[15] C. Liu,[43] H. Liu,[53] Q. Liu,[44] T. Liu,[15] S. Lockwitz,[57] A. Loginov,[57] D. Lucchesi[ff],[40] J. Lueck,[24] P. Lujan,[26] P. Lukens,[15] G. Lungu,[46] J. Lys,[26] R. Lysak[e],[12] R. Madrak,[15] K. Maeshima,[15] P. Maestro[hh],[42] S. Malik,[46] G. Manca[a],[27] A. Manousakis-Katsikakis,[3] F. Margaroli,[47] C. Marino,[24] M. Martínez,[4] P. Mastrandrea,[47] K. Matera,[22] M.E. Mattson,[55] A. Mazzacane,[15] P. Mazzanti,[6] K.S. McFarland,[45] P. McIntyre,[49] R. McNulty[j],[27] A. Mehta,[27] P. Mehtala,[21] C. Mesropian,[46] T. Miao,[15] D. Mietlicki,[32] A. Mitra,[1] H. Miyake,[51] S. Moed,[15] N. Moggi,[6] M.N. Mondragon[m],[15] C.S. Moon,[25] R. Moore,[15] M.J. Morello[ii],[42] J. Morlock,[24] P. Movilla Fernandez,[15] A. Mukherjee,[15] Th. Muller,[24] P. Murat,[15] M. Mussini[ee],[6] J. Nachtman[n],[15] Y. Nagai,[51] J. Naganoma,[54] I. Nakano,[37] A. Napier,[52] J. Nett,[49] C. Neu,[53] M.S. Neubauer,[22] J. Nielsen[d],[26] L. Nodulman,[2] S.Y. Noh,[25] O. Norniella,[22] L. Oakes,[39] S.H. Oh,[14] Y.D. Oh,[25] I. Oksuzian,[53] T. Okusawa,[38] R. Orava,[21] L. Ortolan,[4] S. Pagan Griso[ff],[40] C. Pagliarone,[50] E. Palencia[f],[9] V. Papadimitriou,[15] A.A. Paramonov,[2] J. Patrick,[15] G. Pauletta[kk],[50] M. Paulini,[10] C. Paus,[30] D.E. Pellett,[7] A. Penzo,[50] T.J. Phillips,[14] G. Piacentino,[42] E. Pianori,[41] J. Pilot,[36] K. Pitts,[22] C. Plager,[8] L. Pondrom,[56] S. Poprocki[g],[15] K. Potamianos,[44] F. Prokoshin[cc],[13] A. Pranko,[26] F. Ptohos[h],[17] G. Punzi[gg],[42] A. Rahaman,[43] V. Ramakrishnan,[56] N. Ranjan,[44] I. Redondo,[29] P. Renton,[39] M. Rescigno,[47] T. Riddick,[28] F. Rimondi[ee],[6] L. Ristori[42],[15] A. Robson,[19] T. Rodrigo,[9] T. Rodriguez,[41] E. Rogers,[22] S. Rolli[i],[52] R. Roser,[15] F. Ruffini[hh],[42] A. Ruiz,[9] J. Russ,[10] V. Rusu,[15] A. Safonov,[49] W.K. Sakumoto,[45] Y. Sakurai,[54] L. Santi[kk],[50] K. Sato,[51] V. Saveliev[w],[15] A. Savoy-Navarro[aa],[15] P. Schlabach,[15] A. Schmidt,[24] E.E. Schmidt,[15] T. Schwarz,[15] L. Scodellaro,[9] A. Scribano[hh],[42] F. Scuri,[42] S. Seidel,[35] Y. Seiya,[38] A. Semenov,[13] F. Sforza[hh],[42] S.Z. Shalhout,[7] T. Shears,[27]

P.F. Shepard,[43] M. Shimojima[v],[51] M. Shochet,[11] I. Shreyber-Tecker,[34] A. Simonenko,[13] P. Sinervo,[31] K. Sliwa,[52] J.R. Smith,[7] F.D. Snider,[15] A. Soha,[15] V. Sorin,[4] H. Song,[43] P. Squillacioti[hh],[42] M. Stancari,[15] R. St. Denis,[19] B. Stelzer,[31] O. Stelzer-Chilton,[31] D. Stentz[x],[15] J. Strologas,[35] G.L. Strycker,[32] Y. Sudo,[51] A. Sukhanov,[15] I. Suslov,[13] K. Takemasa,[51] Y. Takeuchi,[51] J. Tang,[11] M. Tecchio,[32] P.K. Teng,[1] J. Thom[g],[15] J. Thome,[10] G.A. Thompson,[22] E. Thomson,[41] D. Toback,[49] S. Tokar,[12] K. Tollefson,[33] T. Tomura,[51] D. Tonelli,[15] S. Torre,[17] D. Torretta,[15] P. Totaro,[40] M. Trovato[ii],[42] F. Ukegawa,[51] S. Uozumi,[25] A. Varganov,[32] F. Vázquez[m],[16] G. Velev,[15] C. Vellidis,[15] M. Vidal,[44] I. Vila,[9] R. Vilar,[9] J. Vizán,[9] M. Vogel,[35] G. Volpi,[17] P. Wagner,[41] R.L. Wagner,[15] T. Wakisaka,[38] R. Wallny,[8] S.M. Wang,[1] A. Warburton,[31] D. Waters,[28] W.C. Wester III,[15] D. Whiteson[b],[41] A.B. Wicklund,[2] E. Wicklund,[15] S. Wilbur,[11] F. Wick,[24] H.H. Williams,[41] J.S. Wilson,[36] P. Wilson,[15] B.L. Winer,[36] P. Wittich[g],[15] S. Wolbers,[15] H. Wolfe,[36] T. Wright,[32] X. Wu,[18] Z. Wu,[5] K. Yamamoto,[38] D. Yamato,[38] T. Yang,[15] U.K. Yang[r],[11] Y.C. Yang,[25] W.-M. Yao,[26] G.P. Yeh,[15] K. Yi[n],[15] J. Yoh,[15] K. Yorita,[54] T. Yoshida[l],[38] G.B. Yu,[14] I. Yu,[25] S.S. Yu,[15] J.C. Yun,[15] A. Zanetti,[50] Y. Zeng,[14] C. Zhou,[14] and S. Zucchelli[ee][6]

(CDF Collaboration)[†]

[1]*Institute of Physics, Academia Sinica, Taipei, Taiwan 11529, Republic of China*

[2]*Argonne National Laboratory, Argonne, Illinois 60439, USA*

[3]*University of Athens, 157 71 Athens, Greece*

[4]*Institut de Fisica d'Altes Energies, ICREA, Universitat Autonoma de Barcelona, E-08193, Bellaterra (Barcelona), Spain*

[5]*Baylor University, Waco, Texas 76798, USA*

[6]*Istituto Nazionale di Fisica Nucleare Bologna,* [ee]*University of Bologna, I-40127 Bologna, Italy*

[7]*University of California, Davis, Davis, California 95616, USA*

[8]*University of California, Los Angeles, Los Angeles, California 90024, USA*

[9]*Instituto de Fisica de Cantabria, CSIC-University of Cantabria, 39005 Santander, Spain*

[10]*Carnegie Mellon University, Pittsburgh, Pennsylvania 15213, USA*

[11]*Enrico Fermi Institute, University of Chicago, Chicago, Illinois 60637, USA*

[12]*Comenius University, 842 48 Bratislava, Slovakia; Institute of Experimental Physics, 040 01 Kosice, Slovakia*

[13]*Joint Institute for Nuclear Research, RU-141980 Dubna, Russia*

[14]*Duke University, Durham, North Carolina 27708, USA*

[15]*Fermi National Accelerator Laboratory, Batavia, Illinois 60510, USA*

[16]*University of Florida, Gainesville, Florida 32611, USA*

[17]*Laboratori Nazionali di Frascati, Istituto Nazionale di Fisica Nucleare, I-00044 Frascati, Italy*

[18]*University of Geneva, CH-1211 Geneva 4, Switzerland*

[19]*Glasgow University, Glasgow G12 8QQ, United Kingdom*

[20]*Harvard University, Cambridge, Massachusetts 02138, USA*

[21]*Division of High Energy Physics, Department of Physics, University of Helsinki and Helsinki Institute of Physics, FIN-00014, Helsinki, Finland*

[22]*University of Illinois, Urbana, Illinois 61801, USA*

[23]*The Johns Hopkins University, Baltimore, Maryland 21218, USA*

[24]*Institut für Experimentelle Kernphysik, Karlsruhe Institute of Technology, D-76131 Karlsruhe, Germany*

[25]*Center for High Energy Physics: Kyungpook National University, Daegu 702-701, Korea; Seoul National University, Seoul 151-742, Korea; Sungkyunkwan University, Suwon 440-746, Korea; Korea Institute of Science and Technology Information, Daejeon 305-806, Korea; Chonnam National University, Gwangju 500-757, Korea; Chonbuk National University, Jeonju 561-756, Korea*

[26]*Ernest Orlando Lawrence Berkeley National Laboratory, Berkeley, California 94720, USA*

[27]*University of Liverpool, Liverpool L69 7ZE, United Kingdom*

[28]*University College London, London WC1E 6BT, United Kingdom*

[29]*Centro de Investigaciones Energeticas Medioambientales y Tecnologicas, E-28040 Madrid, Spain*

[30]*Massachusetts Institute of Technology, Cambridge, Massachusetts 02139, USA*

[31]*Institute of Particle Physics: McGill University, Montréal, Québec, Canada H3A 2T8; Simon Fraser University, Burnaby, British Columbia, Canada V5A 1S6; University of Toronto, Toronto, Ontario, Canada M5S 1A7; and TRIUMF, Vancouver, British Columbia, Canada V6T 2A3*

[32]*University of Michigan, Ann Arbor, Michigan 48109, USA*

[33]*Michigan State University, East Lansing, Michigan 48824, USA*

[34]*Institution for Theoretical and Experimental Physics, ITEP, Moscow 117259, Russia*

[35]*University of New Mexico, Albuquerque, New Mexico 87131, USA*

[36]*The Ohio State University, Columbus, Ohio 43210, USA*

[37]*Okayama University, Okayama 700-8530, Japan*

[38]*Osaka City University, Osaka 588, Japan*

[39]*University of Oxford, Oxford OX1 3RH, United Kingdom*

[40]*Istituto Nazionale di Fisica Nucleare, Sezione di Padova-Trento,* [ff]*University of Padova, I-35131 Padova, Italy*
[41]*University of Pennsylvania, Philadelphia, Pennsylvania 19104, USA*
[42]*Istituto Nazionale di Fisica Nucleare Pisa,* [gg]*University of Pisa,*
[hh]*University of Siena and* [ii]*Scuola Normale Superiore, I-56127 Pisa, Italy*
[43]*University of Pittsburgh, Pittsburgh, Pennsylvania 15260, USA*
[44]*Purdue University, West Lafayette, Indiana 47907, USA*
[45]*University of Rochester, Rochester, New York 14627, USA*
[46]*The Rockefeller University, New York, New York 10065, USA*
[47]*Istituto Nazionale di Fisica Nucleare, Sezione di Roma 1,*
[jj]*Sapienza Università di Roma, I-00185 Roma, Italy*
[48]*Rutgers University, Piscataway, New Jersey 08855, USA*
[49]*Texas A&M University, College Station, Texas 77843, USA*
[50]*Istituto Nazionale di Fisica Nucleare Trieste/Udine,*
*I-34100 Trieste,* [kk]*University of Udine, I-33100 Udine, Italy*
[51]*University of Tsukuba, Tsukuba, Ibaraki 305, Japan*
[52]*Tufts University, Medford, Massachusetts 02155, USA*
[53]*University of Virginia, Charlottesville, Virginia 22906, USA*
[54]*Waseda University, Tokyo 169, Japan*
[55]*Wayne State University, Detroit, Michigan 48201, USA*
[56]*University of Wisconsin, Madison, Wisconsin 53706, USA*
[57]*Yale University, New Haven, Connecticut 06520, USA*

The transverse momentum cross section of $e^+e^-$ pairs in the $Z$-boson mass region of 66–116 GeV/$c^2$ is precisely measured using Run II data corresponding to 2.1 fb$^{-1}$ of integrated luminosity recorded by the Collider Detector at Fermilab. The cross section is compared with two quantum chromodynamic calculations. One is a fixed-order perturbative calculation at $\mathcal{O}(\alpha_s^2)$, and the other combines perturbative predictions at high transverse momentum with the gluon resummation formalism at low transverse momentum. Comparisons of the measurement with calculations show reasonable agreement. The measurement is of sufficient precision to allow refinements in the understanding of the transverse momentum distribution.

* Deceased
† With visitors from: [a]Istituto Nazionale di Fisica Nucleare, Sezione di Cagliari, 09042 Monserrato (Cagliari), Italy, [b]University of CA Irvine, Irvine, CA 92697, USA, [c]University of CA Santa Barbara, Santa Barbara, CA 93106, USA, [d]University of CA Santa Cruz, Santa Cruz, CA 95064, USA, [e]Institute of Physics, Academy of Sciences of the Czech Republic, Czech Republic, [f]CERN, CH-1211 Geneva, Switzerland, [g]Cornell University, Ithaca, NY 14853, USA, [h]University of Cyprus, Nicosia CY-1678, Cyprus, [i]Office of Science, U.S. Department of Energy, Washington, DC 20585, USA, [j]University College Dublin, Dublin 4, Ireland, [k]ETH, 8092 Zurich, Switzerland, [l]University of Fukui, Fukui City, Fukui Prefecture, Japan 910-0017, [m]Universidad Iberoamericana, Mexico D.F., Mexico, [n]University of Iowa, Iowa City, IA 52242, USA, [o]Kinki University, Higashi-Osaka City, Japan 577-8502, [p]Kansas State University, Manhattan, KS 66506, USA, [q]Korea University, Seoul, 136-713, Korea, [r]University of Manchester, Manchester M13 9PL, United Kingdom, [s]Queen Mary, University of London, London, E1 4NS, United Kingdom, [t]University of Melbourne, Victoria 3010, Australia, [u]Muons, Inc., Batavia, IL 60510, USA, [v]Nagasaki Institute of Applied Science, Nagasaki, Japan, [w]National Research Nuclear University, Moscow, Russia, [x]Northwestern University, Evanston, IL 60208, USA, [y]University of Notre Dame, Notre Dame, IN 46556, USA, [z]Universidad de Oviedo, E-33007 Oviedo, Spain, [aa]CNRS-IN2P3, Paris, F-75205 France, [bb]Texas Tech University, Lubbock, TX 79609, USA, [cc]Universidad Tecnica Federico Santa Maria, 110v Valparaiso, Chile, [dd]Yarmouk University, Irbid 211-63, Jordan,

## I. INTRODUCTION

In hadron-hadron collisions at high energies, massive lepton pairs are produced via the Drell–Yan process [1]. In the standard model, colliding partons from the hadrons can interact to form an intermediate $W$ or $\gamma^*/Z$ vector boson that subsequently decays into a lepton pair. Initial state quantum chromodynamic (QCD) radiation from the colliding partons imparts transverse momentum ($P_{\rm T}$) to the boson and produces an accompanying final state jet or jets.

A recent advance in QCD fixed-order perturbative calculations at $\mathcal{O}(\alpha_s^2)$ is the evaluation of the Drell–Yan cross section that is fully exclusive and differential [2]. The exclusive cross section includes both the lepton pair produced via the $W$ or $\gamma^*/Z$ boson intermediate state, and the associated final state partons. It includes finite boson widths, boson-lepton spin correlations, and $\gamma - Z$ interference for the $\gamma^*/Z$ intermediate state.

The QCD calculation of the Drell–Yan-process cross section that is differential in transverse momentum for all values of $P_{\rm T}$ employs a resummation formalism [3] that merges fixed-order calculations with an all-orders sum of large terms from soft and collinear gluon emissions. The dynamics at low $P_{\rm T}$ is factorized into a calculable perturbative form factor and a hadron-level, non-perturbative one that must be measured. The non-perturbative form factor also includes the effect of the intrinsic $P_{\rm T}$ of partons in the hadron. Refinement of the phenomenology needs precise measurements of the transverse momentum differential cross section at low $P_{\rm T}$ from hadron-hadron collisions at various center-of-momentum energies, $\sqrt{s}$.

Previous $p\bar{p}$ measurements at $\sqrt{s} = 0.63$ TeV [4, 5] support the resummation formalism, but with limited statistics. The next $p\bar{p}$ measurements at $\sqrt{s} = 1.8$ TeV [6–9] contributed to the phenomenology at low $P_{\rm T}$ [10]. Recent $p\bar{p}$ measurements at $\sqrt{s} = 1.96$ TeV [11] are precise enough to constrain phenomenological calculations of the Drell–Yan lepton pair $P_{\rm T}$ distribution. Early Large Hadron Collider $pp$ results [12, 13] at $\sqrt{s} = 7$ TeV show agreement with calculations.

In this article, a new and precise measurement of the differential cross section in $P_{\rm T}$ for Drell–Yan lepton pairs from $p\bar{p}$ collisions at $\sqrt{s} = 1.96$ TeV is presented. The specific Drell–Yan process is $p\bar{p} \to e^+e^- + X$, where the $e^+e^-$ pair is produced through an intermediate $\gamma^*/Z$ boson, and $X$ is the hadronic final state associated with the production of the boson. The measurement of the differential cross section is restricted to dielectron pairs within the 66–116 GeV/$c^2$ mass range and is fully corrected to include all boson rapidities, electron phase space, and detector effects. Within this mass range, the dielectron pairs originate mostly from the resonant production and decay of $Z$ bosons.

The cross section, measured using 2.1 fb$^{-1}$ of collisions recorded by the Collider Detector at Fermilab (CDF) during 2002–2007, covers $0 < P_{\rm T} < 350$ GeV/$c$. This range is subdivided into variable-width $P_{\rm T}$ bins. For $P_{\rm T} < 25$ GeV/$c$, the bin width is 0.5 GeV/$c$. The cross section presented for each $P_{\rm T}$ bin is the average bin cross section, $\Delta\sigma/\Delta P_{\rm T}$, where $\Delta\sigma$ is the cross section in a $P_{\rm T}$ bin, and $\Delta P_{\rm T}$ its width.

The $\Delta\sigma/\Delta P_{\rm T}$ measurement depends on the correct modeling of the physics and detector to unfold the effects of the detector acceptance and resolution for the $p\bar{p}$ production of Drell–Yan $e^+e^-$ pairs. The modeling of the physics and detector is data driven. This measurement is an extension of the CDF measurements of the Drell–Yan $e^+e^-$ pair rapidity differential cross section [14], and of the decay-electron angular-distribution coefficients [15] that reflect the polarization state of the intermediate $\gamma^*/Z$ boson produced in $p\bar{p} \to \gamma^*/Z + X$. The $\Delta\sigma/\Delta P_{\rm T}$ measurement uses the same 2.1 fb$^{-1}$ data set and analysis methods developed in those measurements, where both the data and the modeling of the physics and detector are well studied and understood.

Section II provides a brief overview of the QCD calculations of $\Delta\sigma/\Delta P_{\rm T}$ used for comparison with this measurement. Section III provides a summary of CDF and the Tevatron collider at Fermi National Accelerator Laboratory. Section IV reports the selection of electrons and dielectrons for the $\Delta\sigma/\Delta P_{\rm T}$ measurement. Section V details the simulation of the data. Section VI describes the cross section and its measurement. Section VII is the summary.

## II. QCD CALCULATIONS

For the Drell–Yan process, QCD radiation from the colliding partons of the hadrons in the initial state imparts transverse momentum to the lepton pairs. Fixed-order perturbative calculations are expected to become increasingly reliable with larger transverse momentum. However, the Drell–Yan process has two energy scales: the lepton-pair invariant mass and transverse momentum. Difficulties arise in the perturbative calculation when these two scales differ significantly. This is a QCD multi-scale problem. Simpler perturbative QCD calculations usually have one scale, and this scale is often used as the scale in the strong coupling, $\alpha_s$, to control accuracy. In addition, all perturbative QCD calculations have an arbitrary mass factorization scale that separates the hard parton scattering from the soft parton distribution functions (PDFs) of the hadrons. With multiple scales, scale issues can be harder to control and quantify.

At the opposite end corresponding to low transverse momentum, large contributions from soft and collinear gluon emissions begin to dominate and limit the applicability of standard perturbative calculations. The QCD resummation methods are used to overcome this limitation [3]. These resummation methods may be viewed as techniques to control large and unreliable contributions from multiple QCD scales in the low transverse momentum kinematic region.

As neither calculation is expected to be accurate over the entire range of $P_{\rm T}$, it is useful to compare them with measurements. Of interest is the low $P_{\rm T}$ region where the bulk of events is produced. The understanding and proper modeling of QCD at low $P_{\rm T}$ is important for many physics measurements. The Drell–Yan process can be used as a benchmark. The measurement presented here is compared with a recent QCD resummation calculation, RESBOS [10, 16–18], and a state-of-the-art QCD fixed-order $\mathcal{O}(\alpha_s^2)$ calculation (NNLO) of $\Delta\sigma/\Delta P_{\rm T}$, FEWZ2 [2, 19].

The FEWZ2 NNLO calculation is fully exclusive and differential for the final-state leptons and partons, and includes $\gamma^*/Z$ finite decay width and lepton correlation effects. For calculations, the MSTW2008 [20] NNLO nucleon PDFs with their 90% C.L. uncertainties and the default FEWZ2 electroweak parameters of the Fermi coupling constant ($G_\mu$) scheme and fine-structure constant at the $Z$-boson mass ($\alpha_{\rm em}^{-1} = 128$) are used. The QCD factorization and renormalization scales are both set to the $Z$-boson mass. As no significant phase-space restrictions are applied on the final state, except for the 66–116 GeV/$c^2$ dilepton mass range limit, FEWZ2 is used here as an inclusive calculation. The numerical integration accuracy is set to the 1% level.

The RESBOS calculation utilizes the Collins, Soper, and Sterman (CSS) resummation formalism that combines fixed-order perturbative QCD calculations with an all-orders summation of large terms from gluon emissions [3]. The CSS cross section consists of two terms: a $W$ function, which contains the large terms from gluon emissions; and a $Y$ function, which is the fixed-order cross section minus its asymptotic (large gluon emission) terms already in $W$. The $Y$ function becomes important as the magnitude of the $P_{\rm T}$ approachs the lepton-pair invariant mass. After a Fourier transformation from transverse momentum to its conjugate impact-parameter space ($b$), the resummation in the $W$ function is expressed as renormalization group equations [21]. With this formalism, the lepton-pair mass and impact parameter scales are connected by the renormalization group evolution, through which large perturbative terms are reliably controlled. At small $b$, $W$ is evaluated to arbitrary order in the renormalized coupling. At large $b$, hadron level, non-perturbative terms that must be measured become dominant. The methodologies at small and large impact parameters are joined by factorizing $W$ into a perturbative and a non-perturbative form factor. The perturbative form factor uses the impact parameter, $b_* \equiv b/\sqrt{1+(b/b_{\rm max})^2}$, so that it becomes constant in the non-perturbative region.

The CSS gluon resummation $W$ and $Y$ functions should be evaluated to all orders of $\alpha_s$ and then combined to fully describe the physics at all $P_{\rm T}$. However, practical implementations of the CSS gluon resummation formalism evaluate the perturbative $Y$ function and the perturbative part of the resummed $W$ function term to a finite order in $\alpha_s$. Even with a finite order expansion, the CSS gluon resummation formalism provides a good description of the physics at low lepton-pair $P_{\rm T}$. Above a $P_{\rm T}$ value of about the boson mass, the resummed cross section is dominated by the $Y$ function and is close to the pure fixed-order calculation. However, in an intermediate $P_{\rm T}$ zone starting from about half the boson mass, the cancellation between the $W$ and $Y$ functions evaluated at finite order becomes inadequate because of an order mismatch. The $W$ perturbative expansion terms are intrinsically all-orders from the underlying resummation formalism, but the $Y$ terms are strictly finite-order. Within this intermediate $P_{\rm T}$ zone, $W+Y$ loses accuracy and requires compensation in practical implementations of the resummation formalism.

The RESBOS implementations of the $W$ and $Y$ functions are calculated using CTEQ6.6 PDFs [22], and are provided within RESBOS as cross-section tables on a grid of the boson mass, transverse momentum, and rapidity. The RESBOS non-perturbative form factor [10] of the $W$ function for the Drell–Yan process is

$$\exp\left\{\left[-g_1 - g_2\ln\frac{Q}{2Q_0} - g_1 g_3\ln(100x_1x_2)\right]b^2\right\},$$

where $g_1 = 0.21$ GeV$^2$, $g_2 = 0.68$ GeV$^2$, $g_3 = -0.6$, $Q$ is the lepton pair mass, $Q_0 = 1.6$ GeV/$c^2$ (with $b_{\rm max} = 0.5$ GeV$^{-1}$ for $b_*$ in the perturbative form factor), and $x_1x_2 = Q^2/s$. The $g_{1-3}$ are parameters derived from measurements. This form factor describes both low- and high-mass data at various $\sqrt{s}$ from fixed target to colliders. The specific $W$ and $Y$ function cross-section tables used are $W_{321}$ and $Y_k$, respectively, and the numerical integration uncertainties of RESBOS are under 1% and negligible.

The CSS gluon resummation $W$ function has three separate perturbative functions: $A$, $B$, and $C$. In the RESBOS implementation [17] of the $W$ function, $W_{321}$, those functions are evaluated to $\mathcal{O}(\alpha_s^3)$, $\mathcal{O}(\alpha_s^2)$, and $\mathcal{O}(\alpha_s)$, respectively. Its $Y$ function is $\mathcal{O}(\alpha_s^2)$. At large $P_{\rm T}$, RESBOS utilizes both the resummed cross section, $W+Y$, and the $\mathcal{O}(\alpha_s^2)$ fixed-order cross section. The resummed cross section becomes inaccurate in the intermediate transverse momentum region starting from about half of the boson mass because of the intrinsic order mismatch described previously. Therefore, as the $P_{\rm T}$ increases, a matching procedure between the resummed and fixed-order cross section is implemented by RESBOS to provide a reliable prediction over all transverse momentum. This matching is implemented in the $Y_k$ cross-section table[1], and is a non-trivial, phenomenological part of RESBOS. On the other hand, in the transverse momentum region above the order of the boson mass, the RESBOS calculation and its accuracy are similar to the FEWZ2 NNLO inclusive calculation considered here. The RESBOS calculation also includes the full $\gamma^*/Z$ interference effects with a finite decay-width for the $Z$ boson and with lepton correlations. The dominant electroweak corrections are included in the calculation using the effective Born approximation, as done in the LEP electroweak precision measurements.

[1] To reduce the time needed to compute the $Y_k$ cross section table to $\mathcal{O}(\alpha_s^2)$, the computation is implemented by an $\mathcal{O}(\alpha_s)$ calculation with boson mass, rapidity, and transverse momentum dependent NNLO-to-NLO K-factors.

## III. THE EXPERIMENTAL APPARATUS

The CDF II [23] is a general purpose detector at the Fermilab Tevatron Run II $p\bar{p}$ collider whose center-of-momentum energy is 1.96 TeV. The CDF positive $z$-axis is along the proton direction. For particle trajectories, the polar angle $\theta$ is relative to the proton direction and the azimuthal angle $\phi$ is about the beamline axis. The energy and momentum of a particle are denoted as $E$ and $P$, respectively. Their components transverse to the beamline are defined as $E_{\rm T} = E \sin\theta$ and $P_{\rm T} = P \sin\theta$, repectively. The particle rapidity, $y$, is $y = \frac{1}{2}\ln[\,(E + P_{\rm z}c)/(E - P_{\rm z}c)\,]$, where $P_{\rm z}$ is the component of momentum along the $z$-axis. The pseudorapidity of a particle trajectory is $\eta = -\ln\tan(\theta/2)$. Fixed detector coordinates are specified as $(\eta_{\rm det}, \phi)$, where $\eta_{\rm det}$ is the pseudorapidity from the detector center ($z = 0$). Portions of the detector relevant to this analysis are briefly described next.

The central tracker (COT) is a 3.1 m long, open cell drift chamber that extends radially from 0.4 m to 1.3 m. The 2.1 m long silicon tracker surrounds the Tevatron beam pipe and is within the inner radius of the COT. Combined, these two trackers provide efficient, high resolution tracking over $|\eta_{\rm det}| < 1.3$. Both trackers are immersed in a 1.4 T axial magnetic field produced by a superconducting solenoid just beyond the outer radius of the COT.

Outside the solenoid are the central calorimeters, covering $|\eta_{\rm det}| < 1.1$. The forward regions, $1.1 < |\eta_{\rm det}| < 3.6$, are covered by the end-plug calorimeters. All calorimeters are scintillator-based sampling calorimeters read out with phototubes. Both calorimeters are segmented along their depth into electromagnetic (EM) and hadronic (HAD) sections and transversely into projective towers. The EM calorimeter energy resolutions measured in test beams with electrons are $\sigma/E = 14\%/\sqrt{E_{\rm T}}$ for the central calorimeter, and $\sigma/E = 16\%/\sqrt{E} \oplus 1\%$ for the plug calorimeter, where the symbol $\oplus$ is a quadrature sum, and $E_{\rm T}$ and $E$ are in units of GeV. Both the central and plug EM calorimeters have preshower and shower-maximum detectors for electromagnetic shower identification and shower centroid measurements. The combination of the plug shower-maximum detector and silicon tracker provides enhanced tracking coverage to $|\eta_{\rm det}| = 2.8$.

The Fermilab Tevatron collides bunches of protons and anti-protons at a nominal crossing frequency of 2.5 MHz. Over 2002–2007 operations, the instantaneous $p\bar{p}$ collision luminosities at the start of collisions increased over an order of magnitude to $280 \times 10^{30}$ cm$^{-2}$s$^{-1}$. Collision luminosities are continuously measured by the gas Cherenkov counters which are just outside the Tevatron beam pipe and are in the region $3.7 < |\eta_{\rm det}| < 4.7$ [24].

The CDF event trigger system has three tiers, L1, L2, and L3. The L1 trigger is entirely implemented in hardware, is based on trigger primitives, and is synchronous and deadtime-less. Trigger primitives are quantities from the front-end readout used for trigger decisions. The L2 trigger, which processes events selected by the L1 trigger, is asynchronous and is a combination of hardware and software that uses L1 primitives along with additional front-end data. The L3 trigger processes events selected by the L2 trigger and is a speed-optimized version of the CDF offline reconstruction. Track- and EM-objects, which are available at all trigger levels and are refined at each level, form the basis of very efficient trigger paths for the electrons used in this measurement.

## IV. DATA SELECTION

The data set consists of 2.1 fb$^{-1}$ of $p\bar{p}$ collisions at $\sqrt{s} = 1.96$ TeV collected during 2002–2007. Collisions producing massive Drell–Yan dielectron pairs have the following experimental signatures:

- A large fraction of the electrons have high $E_{\rm T}$.
- There are two well-separated electrons of opposite charge.
- The electrons tend to be separated from jets and other particles from the interaction.

These features are used in the selection of events both at the trigger and analysis levels. Electrons in both the central and plug calorimeters are selected.

### A. Triggers

The high $E_{\rm T}$ electrons are selected from generic $p\bar{p}$ collisions by two non-attenuated (full-rate) triggers: the CENTRAL-18, and $Z$-NO-TRACK. Each has well-defined L1, L2, and L3 trigger paths for both physics and trigger efficiency measurements. Independent and dedicated trigger paths are used for the efficiency measurements.

The CENTRAL-18 trigger is the inclusive electron trigger for electrons with $E_{\rm T} > 18$ GeV in the central calorimeter region [23]. A track is required at all trigger levels. Loose criteria applied at each level select candidates that are consistent with an electron showering in the calorimeter, including EM-shower-like lateral shower profile in the EM compartment, EM-shower-like energy leakage in the HAD compartment, and matching between the track and the shower centroid in the EM shower-maximum detector. There is no equivalent inclusive plug electron trigger because the L1 and L2 tracking and the plug calorimeter acceptance do not overlap.

The $Z$-NO-TRACK trigger identifies dielectrons using solely calorimeter information. No tracking information is used. Electron candidates can be in either the central or plug calorimeter region. Both candidates are required to have $E_{\rm T} > 18$ GeV. The only other requirement is that shower energy sharing in the EM and HAD compartments be electron-like. While this trigger is specifically for dielectron candidates that are both in the plug calorimeter region, it accepts the small fraction of dielectron events that fail the CENTRAL-18 trigger.

### B. Electron Selection

To improve the purity of the sample, CDF standard central and plug [23] electron identification requirements are applied. Fiducial requirements are always applied to ensure that the electrons are in well-instrumented regions of CDF where their reconstruction is well understood and predictable. Each electron candidate is required to have an associated track. Having track matching on both electron candidates significantly reduces backgrounds.

The track vertex position along the beamline ($Z_{\rm vtx}$) is restricted to the inner region of CDF: $|Z_{\rm vtx}| < 60$ cm. For 2002–2007 Tevatron operations, 4% of the $p\bar{p}$ luminous region along the beamline is outside this fiducial region. The $p\bar{p}$ collision profile along the beamline is measured by vertexing multiple tracks in minimum-bias events. The multiple track vertexing acceptance is relatively flat within $|Z_{\rm vtx}| \sim 100$ cm.

As electrons in both the central and plug calorimeter regions are used, there are three exclusive Drell–Yan dielectron topologies: CC, CP, and PP, where the C (P) refers to an electron in the central (plug) calorimeter. In the measurement of the $ee$-pair $P_{\rm T}$ distribution, the kinematic region of the $ee$-pair extends over all rapidities, but is restricted to the 66–116 GeV/$c^2$ pair mass range. The kinematic and fiducial regions of acceptance for electrons in the three dielectron topologies are listed below.

1. Central–Central (CC)
   - $E_{\rm T} > 25$ (15) GeV for electron 1 (2)
   - $0.05 < |\eta_{\rm det}| < 1.05$
2. Central–Plug (CP)
   - $E_{\rm T} > 20$ GeV for both electrons
   - Central region: $0.05 < |\eta_{\rm det}| < 1.05$
   - Plug region: $1.2 < |\eta_{\rm det}| < 2.8$
3. Plug–Plug (PP)
   - $E_{\rm T} > 25$ GeV for both electrons
   - $1.2 < |\eta_{\rm det}| < 2.8$

The CC electron $E_{\rm T}$ selection is asymmetric. Electron 1 has the highest $E_{\rm T}$. The asymmetric selection is the result of an optimization based on the decay electron angular distribution measurement [15]. It improves the acceptance in the electron phase space. The PP electron candidates are both required to be in the same end-plug calorimeter, and these pairs extend the rapidity coverage to $|y| \sim 2.9$. At the Tevatron, the kinematic limit for $|y|$ of the $ee$-pair at the $Z$-boson mass is 3.1. Drell–Yan dielectrons in opposing end plug calorimeters have little acceptance, tend to be at low $ee$-pair rapidities, and are overwhelmed by QCD di-jet backgrounds.

As Drell–Yan high-$E_{\rm T}$ leptons are typically produced in isolation, the electron candidates are required to be isolated from other calorimetric activity. The isolation requirement is that the sum of $E_{\rm T}$ over towers within a 0.4 isolation cone in $(\eta, \phi)$ surrounding the electron cluster be under 4 GeV ($E_{\rm iso} < 4$ GeV). The towers of the electron cluster are not included in the sum. While this is a topological selection rather than an electron identification selection, it is included in the electron identification efficiencies.

Electron identification in the central calorimeter region is optimized for electrons of $P_{\rm T} > 10$ GeV/$c$. It utilizes the COT and silicon trackers, the longitudinal and lateral (tower) segmentation of the EM and HAD calorimeter compartments, and the shower-maximum strip detector (CES) within the EM calorimeter. The most discriminating information is provided by the trackers in combination with the CES. An electron candidate must have shower clusters within the EM calorimeter towers and CES that have EM-like lateral shower profiles. A candidate must also have an associated track that extrapolates to the three-dimensional position of the CES shower centroid. The track transverse momentum, $P_{\rm T}$, must be consistent with the associated electron shower $E_{\rm T}$ via an $E/P$ selection when $P_{\rm T} < 50$ GeV/$c$. For both the track matching in the CES and $E/P$ selection, allowances are included for bremsstrahlung energy loss in the tracking volume, which on average is about 20% of a radiation length. The fraction of shower energy in the HAD calorimeter towers behind the EM tower cluster must be consistent with that for electrons ($E_{\rm HAD}/E_{\rm EM}$ requirement). These selections are more restrictive than the ones used in the trigger.

The central electron selection as described has high purity and is called the tight central electron (TCE) selection. Its average selection efficiency is 84%. The track-finding efficiency on the associated tracks is 99%. To improve the selection of central dielectrons, a looser selection, called the loose central electron (LCE) selection, is used on the second electron. The LCE selection does not use transverse shower shape constraints, the $E/P$ constraint, nor track matching in the CES. For track associations, the track need only project into the largest-energy calorimeter tower within the cluster of towers associated with the EM shower. For electron candidates that fail the TCE selection, the LCE selection has an average exclusive efficiency of 76%.

Electron identification in the forward plug calorimeter region also utilizes the COT and silicon trackers, the longitudinal and lateral (tower) segmentation of the EM and HAD calorimeter compartments, and the shower-maximum strip detector (PES) within the EM calorimeter. However, as the plug calorimeter geometry is completely different from the central geometry, the details of the identification requirements differ.

The plate-geometry, end-plug calorimeters have projective towers, but these towers are physically much smaller than the central calorimetry towers. EM showers in the plug calorimeter are clustered into "rectangular" $3 \times 3$ tower clusters in $(\eta, \phi)$ space, with the highest-energy tower in the center. The EM calorimeter energy resolution and lateral shower shapes measured in an electron test beam use $3 \times 3$ shower clustering [25]. The EM preshower detector is the first layer of the EM calorimeter and it is instrumented and read out separately. As there are $\sim 0.7$ radiation lengths of material in front of it, its energy is always included in the EM-cluster shower energy.

An electron in the plug calorimeter, like those in the central region, must also have shower clusters within the EM calorimeter towers and PES that have EM-like lateral shower profiles. The longitudinal $E_{\rm HAD}/E_{\rm EM}$ leakage requirement is more restrictive because of the deeper depth of the EM section and the differing collision conditions in the forward region. The plug selection efficiency without the tracking requirement averages about 84%.

Tracks going into the plug calorimeters have limited geometrical acceptance in the COT for $|\eta_{\rm det}| > 1.3$. The forward tracking coverage of the silicon tracker is exploited with a calorimetry-seeded tracking algorithm called "Phoenix". It is similar to central tracking, where tracks found in the COT are projected into the silicon tracker and hits within a narrow road of the trajectory seed silicon track reconstruction. With the Phoenix algorithm, the track helix in the magnetic field is specified by the position of the $p\bar{p}$ collision vertex, the three-dimensional exit position of the electron into the PES, and a helix curvature. The curvature is derived from the $E_{\rm T}$ of the shower in the EM calorimeter. As the $E_{\rm T}$ provides no information on the particle charge, there are two potential helices, one for each charge. The algorithm projects each helix into the silicon tracker and seeds the silicon track reconstruction. If both projections yield tracks, the higher quality one is selected. The COT is not directly used, but tracks found by the trackers are used to reconstruct the location of the $p\bar{p}$ collision vertex.

The radial extent of the PES, relative to the beamline, is 12–129 cm. Depending on the track vertex location along the beamline ($Z_{\rm vtx}$), a track traverses from 0 to 8 layers of silicon. A Phoenix track is required to have at least three silicon hits. Only plug electrons associated to tracks that traversed at least three silicon layers are accepted. Eighty percent of the tracks traverse four or more silicon layers. Within the plug region, the average Phoenix track acceptance is 94% and within this acceptance zone, the track-finding efficiency is 91%.

The Phoenix algorithm is efficient and results in low background. While the pointing resolution of a Phoenix track is good (1 mrad or better), its path length in the magnetic field at large $|\eta_{\rm det}|$ is small and the helix curvature resolution is poor. Consequently, there is neither a $P_{\rm T}$ nor $E/P$ requirement for plug electron identification.

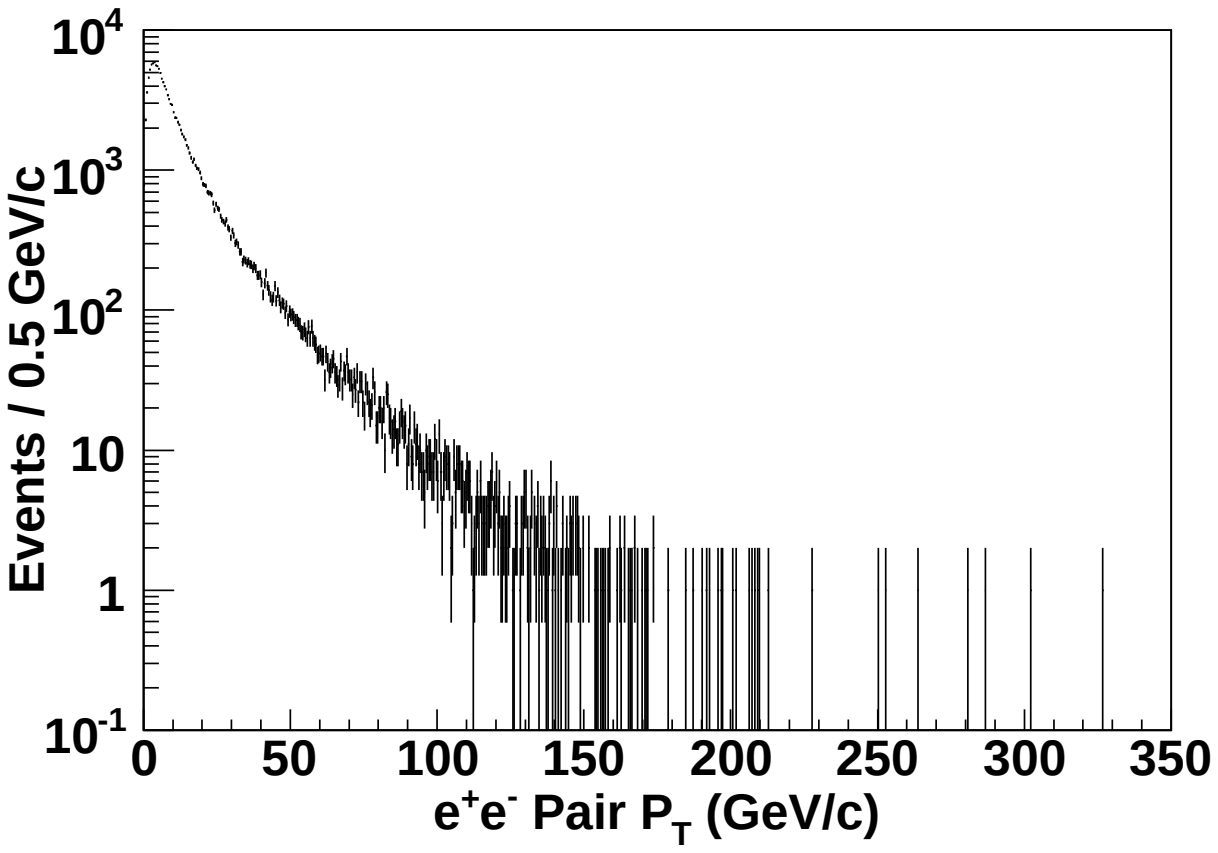

FIG. 1. The raw $ee$-pair $P_{\rm T}$ distribution for all dielectron topologies combined. No corrections or background subtractions are applied. The highest $P_{\rm T}$ is 327 GeV/$c$.

The central region tracking algorithm utilizes hits in the silicon tracker if available. However, the plug Phoenix tracking algorithm requires a fully functional silicon tracker. This silicon requirement reduces the effective integrated luminosity of CP and PP topology dielectrons relative to CC dielectrons by 6%.

### C. Dielectron Selection

Events are required to have a reconstructed dielectron pair mass $66 < M_{ee} < 116$ GeV/$c^2$. For dielectrons of the CC topology, the two tracks are required to have opposite charge. However, for CP and PP topology dielectrons, there is no opposite charge requirement because of the significant charge misidentification on Phoenix tracks at large $|\eta_{\rm det}|$.

The efficiency for the trigger to select events is typically over 99% for dielectrons that pass offline event selections. The CENTRAL-18 trigger has an inefficiency of 3% per single central electron due to track association requirements. The $Z$-NO-TRACK trigger is on average less than 0.5% inefficient for all topologies, and complements the CENTRAL-18 trigger.

### D. Measurement Event Sample

The numbers of events passing all previously described selections in the CC, CP, and PP dielectron topologies are 51 951, 63 752, and 22 469, respectively. Figure 1 shows the raw $ee$-pair $P_{\rm T}$ distribution for these events. The backgrounds are small, and are from QCD or from $WW$, $WZ$, $ZZ$, $t\bar{t}$, $W$+jets, and $Z \to \tau^+\tau^-$ sources with real high-$E_{\rm T}$ electrons. The QCD background is primarily from dijets where a track in a jet fakes an electron or is an electron from a photon conversion. The high-$E_{\rm T}$ electron sources have at least one real electron. The second electron is either a real second electron or a fake one such as in $W$ + jets.

Overall, the background from QCD and non-Drell–Yan high-$E_{\rm T}$ electrons is 0.5%. It is negligible at low pair $P_{\rm T}$, and for $P_{\rm T} > 100$ GeV/$c$, it reaches the 5% level. These backgrounds are subtracted from the $P_{\rm T}$ distribution shown in Fig. 1 for the measurement of $\Delta\sigma/\Delta P_{\rm T}$. Backgrounds are significantly reduced, particularly at large $P_{\rm T}$, by requiring each electron candidate to have an associated track.

The overall QCD background level is 0.3%, and it is under 1% at all $P_{\rm T}$. It is estimated with the data used for the $P_{\rm T}$ measurement using an "isolation extrapolation" procedure. All selection criteria are applied to both electron candidates except the isolation energy ($E_{\rm iso}$) requirement on one electron candidate. Its $E_{\rm iso}$ distribution has a sharp peak at low $E_{\rm iso}$ from Drell–Yan electrons (the signal) and a broad, flat distribution extending to very large $E_{\rm iso}$ from QCD sources (the background). The $E_{\rm iso}$ distribution is fit to a signal plus a background component over the full $E_{\rm iso}$ range, and the background component is extrapolated into the signal region for the QCD background estimate. The signal and background shapes are derived from the unbiased data set used in the measurement, and with selections close to the electron selections to avoid biases. For the background shape event selection, two electron-like candidates are required, but one is selected to be "jet-like" by reversing the selection requirement on its $E_{\rm iso}$ and $E_{\rm HAD}/E_{\rm EM}$ parameters. The other, whose $E_{\rm iso}$ distribution is the background shape, has all electron selection requirements except $E_{\rm iso}$ applied.

The high $E_{\rm T}$ electron backgrounds from $WW$, $WZ$, $ZZ$, $t\bar{t}$, $W$ + jets, and $Z \to \tau^+\tau^-$ are derived from the simulated samples. The overall background level from these sources is 0.2%, but they are the source of the 5% backgrounds for $P_{\rm T} > 100$ GeV/$c$.

Above the $P_{\rm T}$ of 150 GeV/$c$, there are 55 events. The $ee$-pair mass distribution has a clear $Z$-boson mass peak, and within the 66–116 GeV/$c^2$ mass range, there is no indication of unexpected backgrounds. The peak location and width are consistent with expectations.

## V. DATA SIMULATION

The acceptance for Drell–Yan dilepton pairs is obtained using the Monte Carlo physics event generator, PYTHIA 6.214 [26], and the CDF event and detector simulations. PYTHIA generates the hard, leading order (LO) QCD interaction, $q + \bar{q} \to \gamma^*/Z$, simulates initial state QCD radiation via its parton shower algorithms, and generates the decay, $\gamma^*/Z \to l^+l^-$. The CTEQ5L [27] nucleon parton distribution functions (PDFs) are used in the QCD calculations. The underlying event and $\gamma^*/Z$ boson $P_{\rm T}$ parameters are PYTHIA tune AW (i.e., PYTUNE 101, which is a tuning to previous CDF data) [26, 28].

Generated events are processed by the CDF event and detector simulation. The event simulation includes PHOTOS 2.0 [29] which adds final-state quantum electrodynamics (QED) radiation to decay vertices with charged particles, e.g. $\gamma^*/Z \to e^+e^-$. The time-dependent beam and detector conditions for data runs recorded and used for physics analyses are simulated. The beam conditions simulated are the $p$ and $\bar{p}$ beamline parameters, the $p\bar{p}$ luminous region profile, and the instantaneous and integrated luminosities per run. The detector conditions simulated are detector component calibrations, which include channel gains and malfunctions. Thus, the simulated events parallel the recorded data, and are reconstructed, selected, and analyzed as the data.

The $\Delta\sigma/\Delta P_{\rm T}$ measurement is data driven and depends on the correct modeling of both the physics and the detector. The procedure involves the measurement and tuning of the underlying kinematics and detector parameters that make the simulated, reconstructed event distributions match the actual data as precisely as possible. This is a bootstrap process that iterates if necessary for the required precision. The default simulation does not reproduce the data at the precision required. The following subsections describe the model tunings.

### A. Physics Simulation

The Drell–Yan dilepton production is described by

$$\frac{d^4\sigma}{dM^2\,dy\,dP_{\rm T}d\Omega} = \frac{d^3\sigma}{dM^2\,dy\,dP_{\rm T}}\;\frac{dN}{d\Omega}\,,$$

where $d^3\sigma/dM^2\,dy\,dP_{\rm T}$ is the unpolarized $\gamma^*/Z$ boson production cross section at the resonance mass $M$ with subsequent decay to $e^+e^-$, and $dN/d\Omega$ the electron angular distribution of the $\gamma^*/Z \to e^+e^-$ decay in a boson rest frame. For this measurement, the single differential distributions $d\sigma/dy$ and $d\sigma/dP_{\rm T}$, and the electron angular distribution, are tuned to the data. The $y$ distribution tuning for $\gamma^*/Z$ production is from the $d\sigma/dy$ measurement [14]. The tuning of the electron angular distribution is briefly reviewed next. The tuning of $d\sigma/dP_{\rm T}$ is specific to this analysis, and is presented last.

The PYTHIA parton showering starts with the $q\bar{q} \to \gamma^*/Z$ annihilation vertex at the end of the shower chain then evolves the shower backwards in time to an initiating $q\bar{q}$ or $qg$ state. The Compton production process cannot be fully simulated. While its gluon splitting subprocess is simulated, the gluon fusion subprocess, $qg \to q^* \to q + \gamma^*/Z$, cannot be simulated from the annihilation vertex at the end of the shower chain. The gluon fusion production rate is compensated in the shower, but there is no compensation to the boson polarization states affected by this subprocess. The boson polarization affects the decay electron angular distribution.

The decay electron angular distribution is analyzed in the Collins–Soper (CS) rest frame [30] of the $e^+e^-$ pair. The CS frame is reached from the laboratory frame via a Lorentz boost along the lab $z$-axis into a frame where the $z$-component of the pair momentum is zero, followed by a boost along the $P_{\rm T}$ of the pair. At $P_{\rm T} = 0$, the CS and laboratory coordinate frames are the same. Within the CS frame, the $z$-axis for the polar angle is the angular bisector between the proton direction and the negative of the anti-proton direction. The $x$-axis is the direction of the $P_{\rm T}$. The polar and azimuthal angles of the $e^-$ in the rest frame are denoted as $\vartheta$ and $\varphi$, respectively.

The general structure of the Drell–Yan decay lepton angular distribution in a boson rest frame consists of nine helicity cross sections governed by the polarization state of the vector boson [31],

$$\begin{aligned}\frac{16\pi}{3}\,\frac{dN}{d\Omega} = \; & (1+\cos^2\vartheta) + \\ & A_0\,\frac{1}{2}\,(1-3\cos^2\vartheta) + \\ & A_1\;\sin 2\vartheta\cos\varphi + \\ & A_2\,\frac{1}{2}\;\sin^2\vartheta\cos 2\varphi + \\ & A_3\;\sin\vartheta\cos\varphi + \\ & A_4\;\cos\vartheta + \\ & A_5\;\sin^2\vartheta\sin 2\varphi + \\ & A_6\;\sin 2\vartheta\sin\varphi + \\ & A_7\;\sin\vartheta\sin\varphi\,.\end{aligned}$$

The $A_{0-7}$ coefficients are cross section ratios, and are functions of the boson kinematics. They are zero at $P_{\rm T} = 0$, except for the electroweak part of $A_4$ responsible for the forward-backward $e^-$ asymmetry in $\cos\vartheta$. The $A_{5-7}$ coefficients appear at $\mathcal{O}(\alpha_s^2)$ and are small in the CS frame. The decay-electron angular-distribution analysis [15] in the CS frame measures the large and accessible decay electron angular coefficients, $A_0$, $A_2$, $A_3$, and $A_4$, as functions of $P_{\rm T}$. These measurements are incorporated into the modeling of $\gamma^*/Z \to e^+e^-$ decays.

The generator-level $P_{\rm T}$ distribution is adjusted, bin by bin, so that the shape of the reconstruction-level, simulated $P_{\rm T}$ distribution is the same as in the data. The method uses the data-to-simulation ratio of the number of reconstructed events in $P_{\rm T}$ bins as an iterative adjustment estimator for the generator level $P_{\rm T}$ bins. Successive iterations unfold the smearing of events across $P_{\rm T}$ bins. Figure 2 is the generator-level $P_{\rm T}$ correction function that makes the data-to-simulation ratio uniform. Statistical fluctuations in the ratio are smoothed out. The $\Delta N/\Delta P_{\rm T}$ correction is the measurement of the shape of $d\sigma/dP_{\rm T}$ that is used in the physics model.

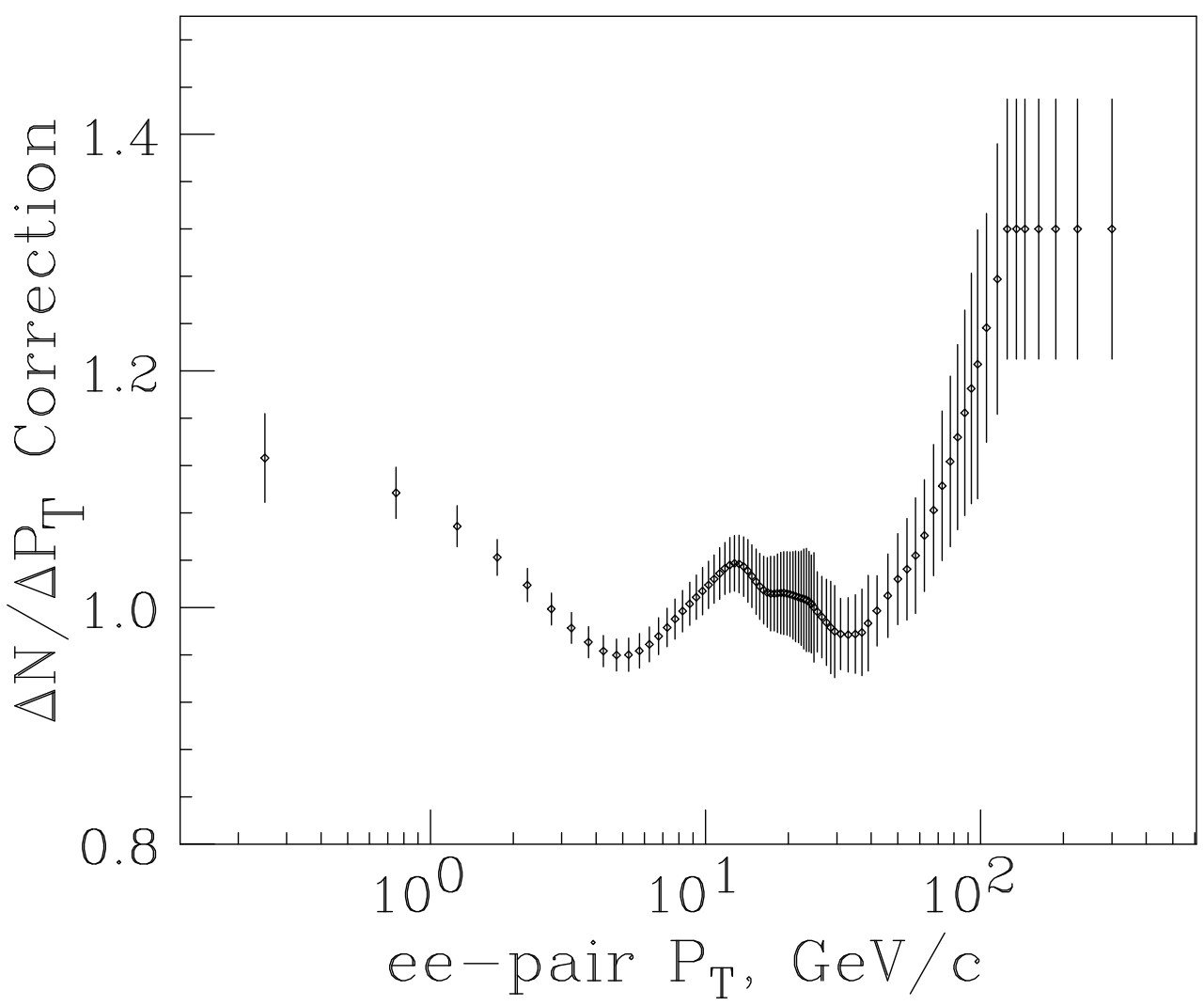

FIG. 2. The $P_{\rm T}$ correction function applied to the generator level $\Delta N/\Delta P_{\rm T}$ distribution that makes flat the ratio of the observed data to the simulated data. The points are at the center of the $P_{\rm T}$ bins. For the low-statistics $P_{\rm T} > 120$ GeV/$c$ region, an average correction is used.

### B. Detector Simulation

The simulation is used to calculate the combined detector acceptance ($A$) and selection efficiency ($\epsilon$) as a function of kinematic variables for Drell–Yan dielectrons. The combined acceptance and efficiency convolution is denoted as $A\otimes\epsilon$. Single-electron selection efficiencies are measured and incorporated into the simulation as event-weight scale factors. The scale factors are ratios of the measured efficiencies of the data to the simulated data.

The electron-trigger efficiencies have an $E_{\rm T}$ (calorimetry) and $\eta_{\rm det}$ (tracking) dependence that are measured and incorporated into the simulation. The electron-identification efficiencies are measured as a function of $\eta_{\rm det}$ for both central and plug region electrons. Plug region efficiencies are measured separately for CP and PP topology dielectrons due to their different environments. Plug-electron efficiencies have a clear time dependence due to the increasing instantantaneous luminosities delivered by the Tevatron. This dependence is incorporated into the simulation. Luminosity effects are measured using the number of $p\bar{p}$ vertices reconstructed by the trackers per event.

A precise model of the calorimeter response in the simulation is important for the calculation of $A\otimes\epsilon$. Electron kinematics are derived from a three-momentum that uses the electron energy measured in the calorimeters for the momentum magnitude and the associated track for the direction. The simulated electron energy scale calibration and resolution versus $\eta_{\rm det}$ are tuned using the electron $E_{\rm T}$ distribution. The default scale and resolution per $\eta_{\rm det}$ bin are adjusted so that the electron $E_{\rm T}$ distribution reconstructed in simulation matches that of the data. Only the constant term in the energy resolution is adjusted. Since the default simulation parametrization of the energy resolution can already have a constant term, the resolution adjustment is done with an additional constant term $c_2$,

$$\frac{\sigma}{E} = \frac{c_0}{\sqrt{E}} \oplus c_1 \oplus c_2 \ ,$$

where $\sigma$ is the energy resolution, $E$ is the energy, $c_0$ and $c_1$ are the default parameters of Section III, and the $\oplus$ denotes combination in quadrature. The tuned values of $c_2$ on average are 0.9% and 2.3% for the central and plug calorimeters, respectively. The steeply rising and falling parts of the electron $E_{\mathrm{T}}$ distribution dominate the constraints. The three dielectron topologies, CC, CP, and PP, provide multiple and independent central and plug electron $E_{\mathrm{T}}$ samples. The $\eta_{\mathrm{det}}$-dependent $E_{\mathrm{T}}$ distributions of each topology are calibrated independently. After the $\eta_{\mathrm{det}}$-dependent parameters are determined, the separate CC, CP, and PP dielectron mass distributions are used to set an overall global scale and resolution adjustment for central and plug electrons.

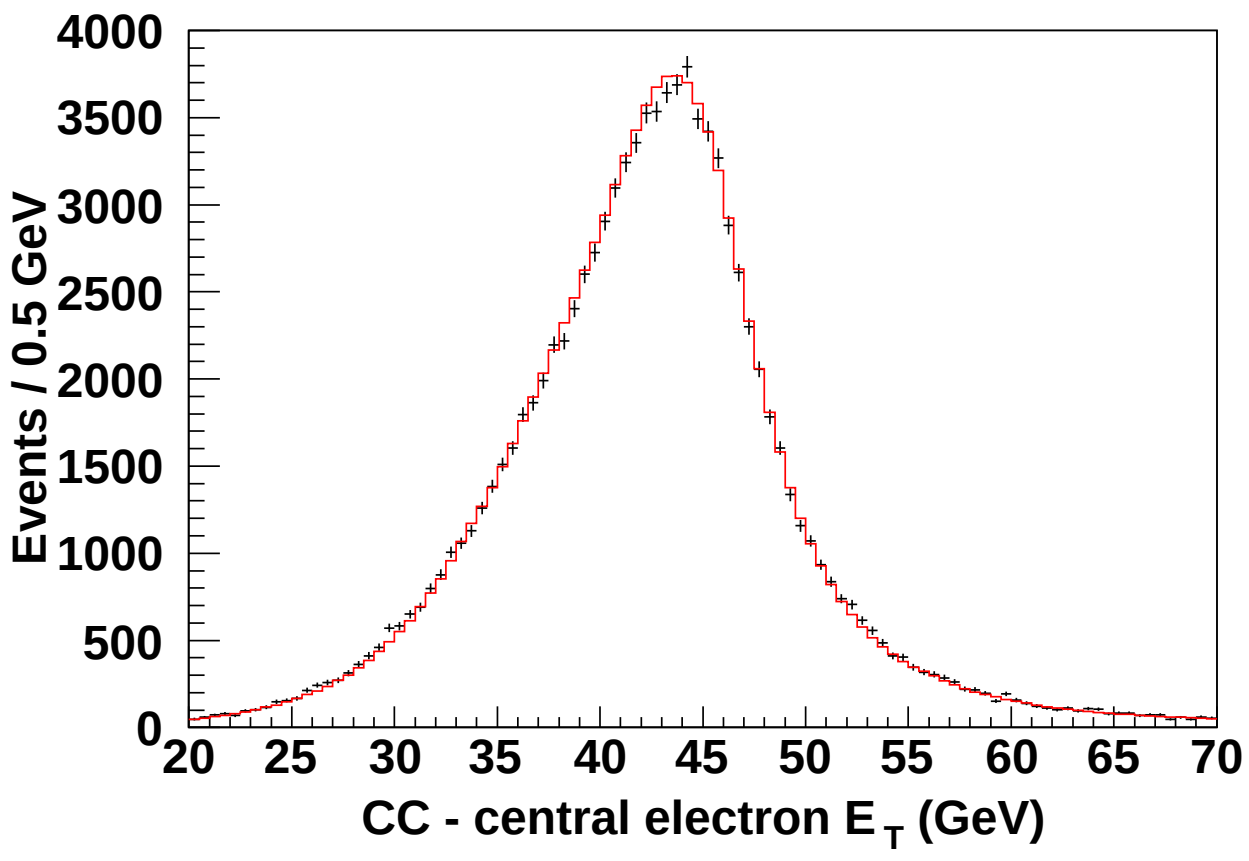

FIG. 3. The overall CC topology central electron $E_{\mathrm{T}}$ distribution. The crosses are the data and the histogram is the simulated data.

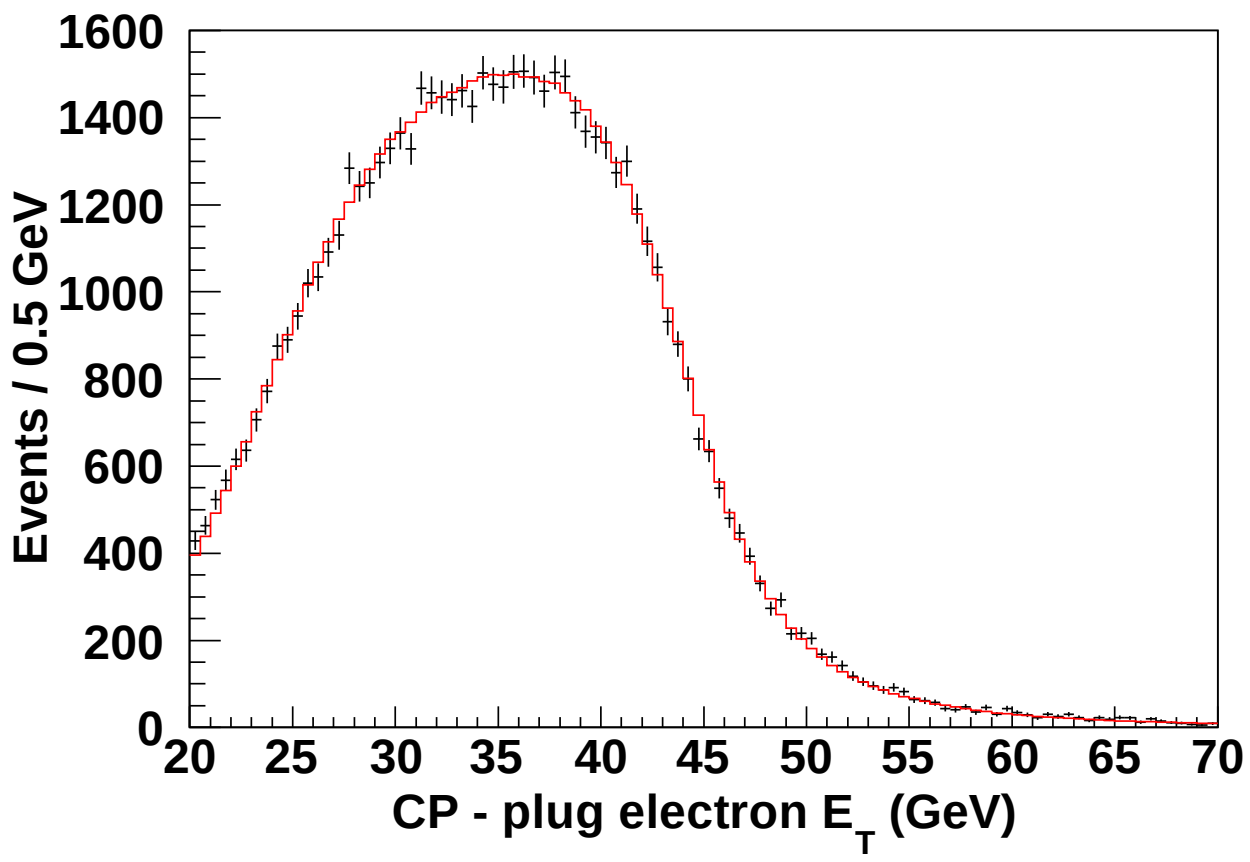

FIG. 4. The overall CP-topology plug electron $E_{\mathrm{T}}$ distribution. The crosses are the data and the histogram is the simulated data.

The simulation is compared to data using histogrammed electron $E_{\mathrm{T}}$ and $ee$-pair mass distributions. Since the backgrounds are small, they are ignored. The comparison statistic is the $\chi^2$ between the simulation and data. The event count of the simulated data is normalized to that of the data, and only statistical uncertainties are used in the calculation.

The $\eta_{\mathrm{det}}$-dependent calorimeter response tunings provide a good match between the simulated-data and data. Figure 3 shows the $E_{\mathrm{T}}$ distribution of CC-topology central electrons. The corresponding plot for PP-topology plug electrons is similar in shape except that the width of the $E_{\mathrm{T}}$ "peak" is slightly narrower. Figure 4 shows the $E_{\mathrm{T}}$ distribution of CP-topology plug electrons. The corresponding plot for CP-topology central electrons is very similar. A $\chi^2$ test is used to evaluate the compatibility between the simulation and data. For CC-central, CP-central, CP-plug, and PP-plug electrons, the $\chi^2$ values are 117, 100, 87, and 135, respectively, for 100 bins (90 bins for PP).

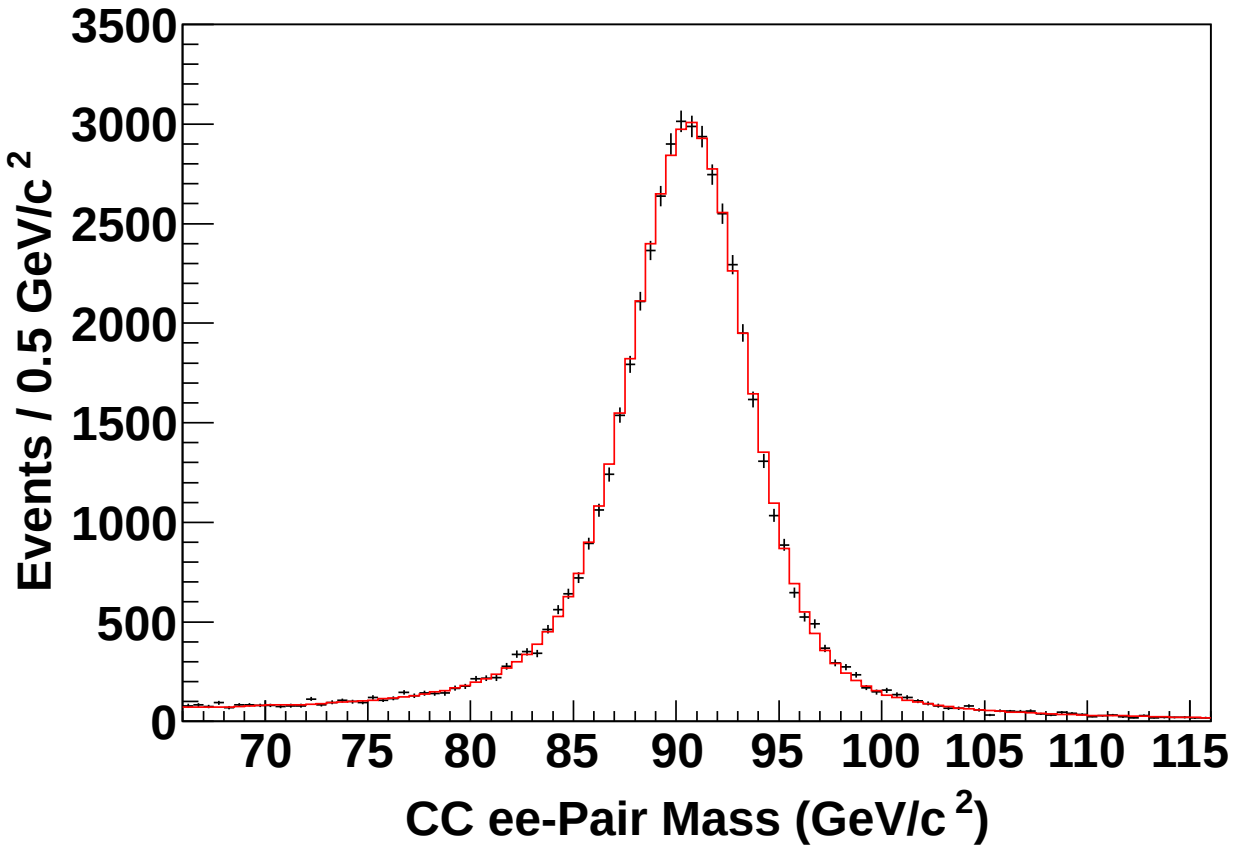

FIG. 5. The overall CC topology $ee$-pair mass distribution. The crosses are the data and the histogram is the simulated data.

Figure 5 shows the CC-topology $ee$-pair mass distribution. The $ee$-pair mass distributions for the CP and PP topologies are similar. The simulated-data to data $\chi^2$ for the CC-, CP-, and PP-topology $ee$-pair mass distributions are 107, 123, and 114, respectively, for 100 bins. The sharp and narrow $Z$-peaks provide significant constraints on the the global energy scale and resolution parameters.

## VI. THE CROSS SECTION

The differential cross section in $P_{\mathrm{T}}$ is the average cross section in a $P_{\mathrm{T}}$ bin, or $\Delta\sigma/\Delta P_{\mathrm{T}}$, where $\Delta\sigma$ is the inte-grated cross section in a bin. The $\Delta\sigma$ is defined as:

$$\Delta\sigma = \frac{N}{\mathcal{L}\ \ A \otimes \epsilon}$$

where $N$ is the background subtracted event count, $\mathcal{L}$ is the effective integrated luminosity, and $A \otimes \epsilon$ is the combined acceptance and efficiency. The effective luminosity, $\mathcal{L}$, is 2057 pb$^{-1}$, and it includes the acceptance of the $|Z_{\rm vtx}| < 60$ cm fiducial restriction. The details of the measurement and its uncertainties are presented next.

### A. Acceptance and Efficiency Unfolding

The combined acceptance and efficiency, $A \otimes \epsilon$, is calculated using the simulation to convolve individual electron $\eta_{\rm det}$ acceptances and efficiencies into an $ee$-pair $P_{\rm T}$ quantity. The value of $A \otimes \epsilon$ ranges from 0.22 at $P_{\rm T} \simeq 0.2$ GeV/$c$ to 0.30 at $P_{\rm T} \simeq 200$ GeV/$c$. As $P_{\rm T}$ increases, the $ee$-pair rapidity becomes more central, the electron $E_{\rm T}$ becomes larger, and the acceptance slowly increases.

The smearing of the observed $P_{\rm T}$ away from the generator ($\gamma^*/Z$) level value is significant relative to the bin size at low $P_{\rm T}$: It has an rms width of about 2.2 GeV/$c$ and is non-Gaussian. Detector resolution and QED radiation from the $\gamma^*/Z \to e^+e^-$ vertex induce distortions to the reconstructed $ee$-pair mass and $P_{\rm T}$ distributions. In addition, they induce a broad enhancement in the $A \otimes \epsilon$ function. It rises from 0.22 at $P_{\rm T} \simeq 0.2$ GeV/$c$ to a broad maximum of 0.28 around a $P_{\rm T}$ of 8 GeV/$c$, then decreases to 0.24 at $P_{\rm T} \sim 30$ GeV/$c$ before increasing again at larger $P_{\rm T}$ due to the increased acceptance.

When $A \otimes \epsilon$ is used to calculate cross sections, it unfolds the effects of smearing. The $(A \otimes \epsilon)^{-1}$ correction is applied bin-by-bin and consists logically of two steps. The first step is a scaling correction on the number of reconstructed and selected events. This scales (unfolds) the number of events reconstructed in a $P_{\rm T}$ bin into the number of reconstructed events produced in the bin. The simulation provides an average scaling factor. The second step is a standard detector acceptance correction on this scaled (unfolded) event count.

For the cross section uncertainty evaluation, more information on event production and migration among the $P_{\rm T}$ bins is required. The number of events produced in each bin have statistical fluctuations. With smearing, there is event migration among the bins, and this migration is also subject to statistical fluctuations. At low $P_{\rm T}$, event migration between bins is large. If these migrations are unaccounted, the cross section uncertainty will be significantly underestimated. As the event migration between bins is not measured, these migrations are estimated with the simulation.

### B. Unfolding Uncertainty Model

Comparisons of fully-corrected cross section measurements with theoretical cross sections are not straightforward. Where detector smearing is significant, there are significant uncertainty correlations among the $P_{\rm T}$ bins due to the event migrations among the bins. The simulation behind the scaling correction accounts for these migrations. The scaling correction uncertainty has both statistical and systematic biases. The systematic bias is from the residual simulation model bias on the bin scaling factor. This bias has been mitigated by the model tuning described in Section V B. The sources of statistical uncertainty from event migration for the scaling correction are discussed, and a model of per-measurement (per-single-experiment) fluctuations for the uncertainty that uses the simulation is specified.

Within the context of the simulation, information about the event migration of reconstructed events among $P_{\rm T}$ bins is in its transfer matrix, $\bar{n}_{lk}$, where $\bar{n}_{lk}$ is the expectation value of the number of events produced in bin $k$ that migrate into bin $l$. The expectation value of a quantity is denoted with an overbar, e.g. $\bar{n}$. All expectation values are normalized to the integrated luminosity of the data. The number of events that do not migrate out of a bin is denoted by $\bar{n}_g$. The number of events that migrate out and in are denoted by $\bar{n}_o$ and $\bar{n}_i$, respectively. In terms of the transfer matrix, $\bar{n}_{lk}$, the $\bar{n}_g$, $\bar{n}_o$, and $\bar{n}_i$ for $P_{\rm T}$ bin $m$ are, respectively, $\bar{n}_{mm}$, the sum of $\bar{n}_{lm}$ over the migration index $l$ excluding bin $m$, and the sum of $\bar{n}_{mk}$ over the production index $k$ excluding bin $m$. The per-measurement statistical fluctuation of a quantity from its expectation value is denoted by $\delta$ followed by the quantity, e.g., $\delta n = n - \bar{n}$. An ensemble variance is denoted by $\delta^2$, e.g., for Poisson statistics, $\delta^2 n = \bar{n}$, and if $c$ is a constant, $\delta^2 cn = c^2\,\bar{n}$.

The scaling correction factor is $\bar{\rho} \equiv \bar{N}_g/\bar{N}_r$, where $\bar{N}_g = \bar{n}_g + \bar{n}_o$ is the expectation on the number of events produced in a bin, and $\bar{N}_r = \bar{n}_g + \bar{n}_i$ is the number of events reconstructed in a bin. Any residual model systematic bias is in $\bar{\rho}$. For a given measurement, the number of events produced and reconstructed in a bin are $N_g = n_g + n_o$ and $N_r = n_g + n_i$, respectively. The scaling correction estimate for $N_g$ is $\bar{\rho}\,N_r$. The difference between the scaling correction estimator $\bar{\rho}\,N_r$ and its target $N_g$ gives a bias between them, $B = \bar{\rho}\,N_r - N_g$. If there are no target fluctuations ($N_g = \bar{N}_g$), $B$ is the statistical bias of the estimator. With target fluctuations, there are two statistical biases, $\bar{\rho}\,N_r - \bar{\rho}\,\bar{N}_r$ ($= \delta\bar{\rho}\,N_r$) and $N_g - \bar{N}_g$ ($= \delta N_g$), and $B$ is their difference.

With no smearing, the estimator and target, along with their fluctuations, are identical, so $B = 0$ and the statistical uncertainty of the scaling correction is just that of the estimator. With smearing, the estimator and target fluctuations are not fully correlated, so $B \neq 0$ and the scaling correction statistical uncertainty is from a combination of estimator and target statistical fluctuations. The estimator ($\bar{\rho}\,N_r$) and target ($N_g$) have three sta-tistically independent elements: $n_g$, $n_o$, and $n_i$. As $n_g$ is part of both the estimator and target, the common overlap must be removed to avoid double counting. The total per-measurement fluctuation for the scaling correction, denoted as $\delta N'_g$, is defined as the sum of fluctuations ($\delta n = n - \bar n$) from the estimator and the target minus their common term, $\delta n_g$:

$$\begin{aligned}\delta N'_g &= \delta\,\bar\rho\,N_r + \delta N_g - \delta n_g \\ &= \delta\,\bar\rho\,(n_g + n_i) + \delta(n_g + n_o) - \delta n_g \\ &= \delta\,\bar\rho\,(n_g + n_i) + \delta n_o\ .\end{aligned}$$

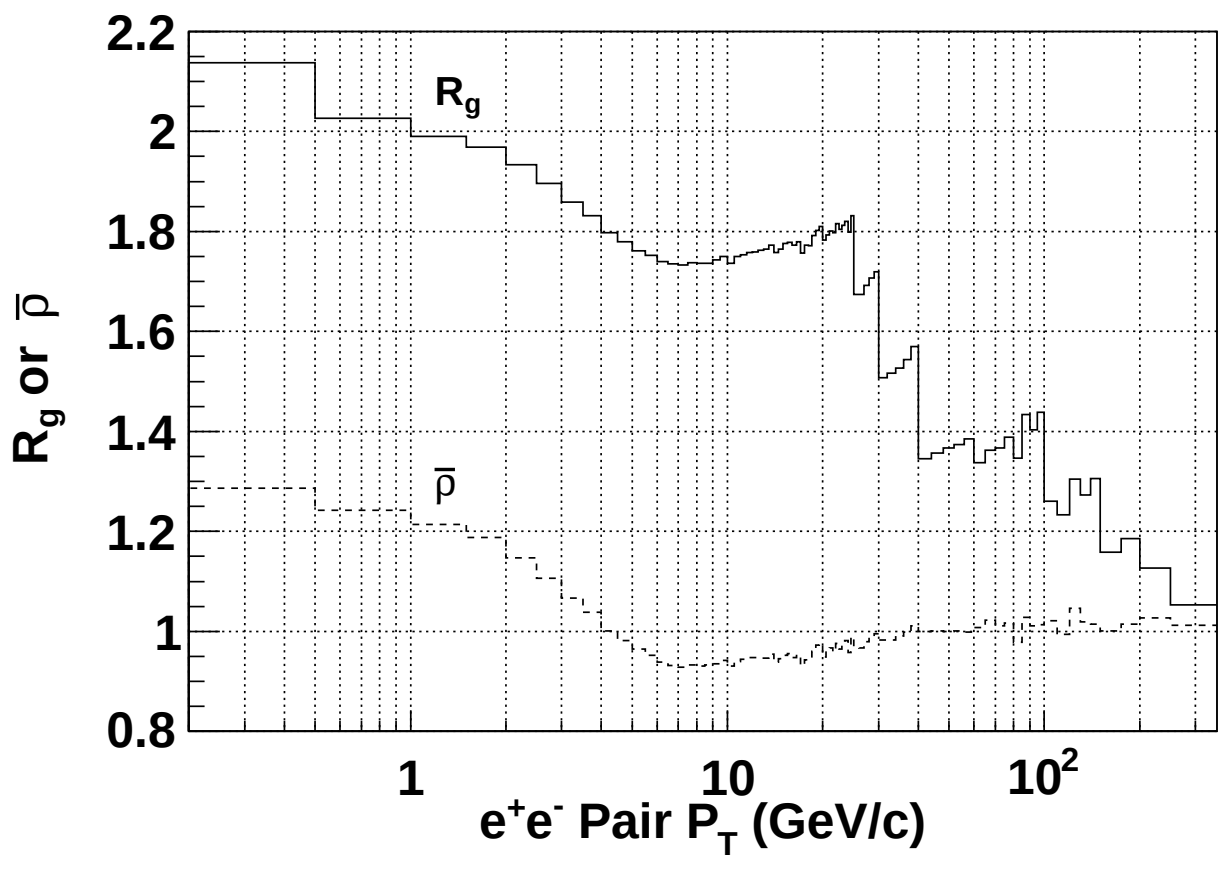

FIG. 6. $R_g(P_\mathrm{T})$ and $\bar\rho(P_\mathrm{T})$. The solid histogram is $R_g$, the bin variance of the uncertainty model relative to the variance of the produced events. The abrupt drops are where the bin size changes. The lower, dashed histogram is $\bar\rho$.

For Poisson statistics, the $P_\mathrm{T}$ bin ensemble variance is:

$$\begin{aligned}\delta^2 N'_g &= \bar\rho^2(\bar n_g + \bar n_i) + \bar n_o \\ &= \bar\rho \bar N_g + \bar n_o\ ,\end{aligned}$$

where $\bar N_g = \bar\rho \bar N_r = \bar\rho(\bar n_g + \bar n_i)$ is used. The covariance from the $\delta n_o$ and $\delta\,\bar\rho\,n_i$ terms between bins $k$ and $l$ is $\bar\rho_k \bar n_{kl} + \bar\rho_l \bar n_{lk}$.

The ratio, $R_g$, of $\delta^2 N'_g$ to $\delta^2 N_g$ ($= \bar N_g$) is the variance of the model relative to the variance of only the produced events. Figure 6 shows both the ratio and the scaling correction factor as functions of $P_\mathrm{T}$. In the low $P_\mathrm{T}$ bins, $\bar n_o$ and $\bar n_i$ are separately much larger than $\bar n_g$. Their effects are significant as $R_g = \bar\rho + \bar n_o/\bar N_g$.

For the uncertainty evaluations, the cross section is rewritten as $\Delta\sigma = \bar\rho N_r/(\mathcal{L}\,A')$, where $A' \equiv \bar\rho\ A\otimes\epsilon$. The uncertainty on $\mathcal{L}$ is systematic and is considered separately. Thus, the fractional uncertainty on $\Delta\sigma$ is a combination of the fractional uncertainty of $\bar\rho N_r$ and $A'$. The fractional uncertainty of $\bar\rho N_r$ is defined as the uncertainty of $\bar\rho N_r$ from the model ($\delta N'_g$) divided by $\bar\rho \bar N_r$ ($= \bar N_g$). The correlation of these fractional uncertainties between $P_\mathrm{T}$ bins $l$ and $k$ is given by the fractional covariance matrix: $\bar V_{lk}/(\bar N_{gl}\,\bar N_{gk})$, where $\bar V_{lk}$ is the covariance matrix of the model, and $\bar N_{gl}$ and $\bar N_{gk}$ are the $\bar N_g$ of bin $l$ and $k$, respectively. The small acceptance fractional uncertainties are added in quadrature to the diagonal part of the fractional covariance matrix. The measured cross sections are used to convert the unitless fractional matrix into units of cross section squared, and this matrix is used to propagate uncertainties for the total cross section measurement and for the comparison of a prediction with the measured cross section.

### C. Systematic Uncertainties

The largest source of uncertainty is the effective integrated luminosity, $\mathcal{L}$. It has an overall uncertainty of 5.8% that consists of a 4% uncertainty of the acceptance of the gas Cherenkov luminosity detector [24] to $p\bar p$ inelastic collisions and a 4.2% measurement uncertainty. It is common to all $P_\mathrm{T}$ bins and not explicitly included. The acceptance uncertainty is primarily from the uncertainty in the beamline and detector geometry (material), and from the uncertainty in the model of the inelastic cross section. The inelastic cross section model contributes 2% to the acceptance uncertainty. The measurement uncertainty contains the uncertainty of the absolute $p\bar p$ inelastic cross section.

The uncertainty on $A\otimes\epsilon$ has a component from the input electron efficiency measurements which depend on $\eta_\mathrm{det}$ and instantaneous luminosity. The simulation is used to propagate these electron measurement uncertainties into an uncertainty for the $ee$-pair $P_\mathrm{T}$ and to include correlations of the same measurements. The calculated uncertainty is uniform and amounts to about 1% over $0 < P_\mathrm{T} < 20$ GeV/$c$. It slowly decreases at higher $P_\mathrm{T}$. A large fraction of the uncertainty is due to plug electron measurement uncertainties. The fractional uncertainty decreases with $P_\mathrm{T}$ because the fraction of plug events decreases. Because the same measurements are used on all $P_\mathrm{T}$ bins, the uncertainty is treated as fully correlated across bins.

The calorimeter response modeling uncertainty analysis is limited by the statistical precision of the simulated data. At the peak of the $P_\mathrm{T}$ distribution, the statistical uncertainty is 0.3%. The variations on the central and plug calorimeter global energy scale and resolutions tunings allowed by the data propagate into changes of $A\otimes\epsilon$ that are no larger than its statistical uncertainty. These changes are not independent.

### D. Results

The Drell–Yan $\Delta\sigma/\Delta P_\mathrm{T}$ for $e^+e^-$ pairs in the $Z$-boson mass region of $66 - 116$ GeV/$c^2$ is shown in Fig. 7 and tabulated in Table I. The total cross section from the numerical integration of the cross section in each $P_\mathrm{T}$ bin is $256.1 \pm 1.3 \pm 2.6$ pb, where the first uncertainty is statistical and the second is the systematic uncertainty due to electron efficiency measurements. The 5.8% integrated luminosity uncertainty of 14.9 pb is not included.

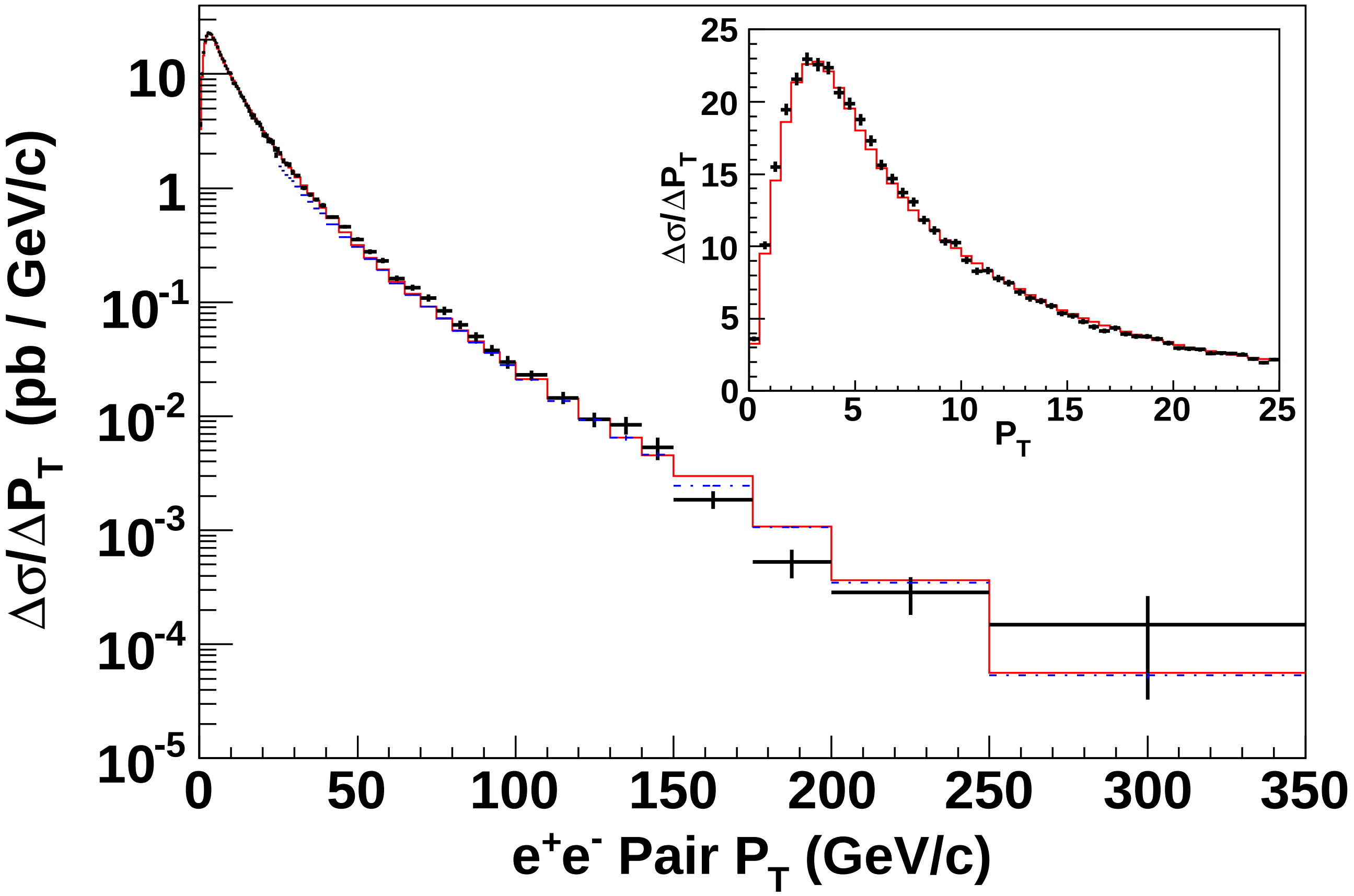

FIG. 7. The $\Delta\sigma/\Delta P_{\mathrm{T}}$ cross section versus $P_{\mathrm{T}}$. Cross section values are plotted at the bin center. The horizontal bars represent the bin extent and the vertical bars are the cross section uncertainties. The solid (black) crosses are the data and all uncertainties except the integrated luminosity uncertainty are combined and plotted. The solid (red) histogram is the RESBOS calculation. The dot-dash (blue) bars of the $P_{\mathrm{T}} > 25$ GeV/$c$ region are the FEWZ2 calculation. For the calculations, only numerical uncertainties are included but they are too small to be visible. The inset is the $P_{\mathrm{T}} < 25$ GeV/$c$ region with a linear ordinate scale.

Figure 7 shows that the RESBOS prediction has a general agreement with the data over the full range of $P_{\mathrm{T}}$. The RESBOS total cross section from the numerical integration of its cross section in each $P_{\mathrm{T}}$ bin is 254 pb. Figure 8 shows the ratio of the measured cross section to the RESBOS prediction in the lower $P_{\mathrm{T}}$ region.

The detector smearing correlates neighboring $P_{\mathrm{T}}$ bin uncertainties that are estimated with the model specified in Section VI A. For the low $P_{\mathrm{T}}$ bins, the correlations spread across many bins but for $P_{\mathrm{T}} > 40$ GeV/$c$, the correlations are predominantly between nearest neighbors. The cross section covariance matrix eigenvalues and eigenvectors are used for the $\chi^2$ comparison between the data and RESBOS. The eigenvalues are the measurement uncertainties (variances) of the associated eigenvector. Measurement uncertainties between eigenvectors are uncorrelated. As an eigenvector corresponds to many $P_{\mathrm{T}}$ bins because of smearing, its most probable $P_{\mathrm{T}}$ bin is used for its association to a $P_{\rm T}$ bin. The mapping of eigenvectors to $P_{\rm T}$ bins is described next.

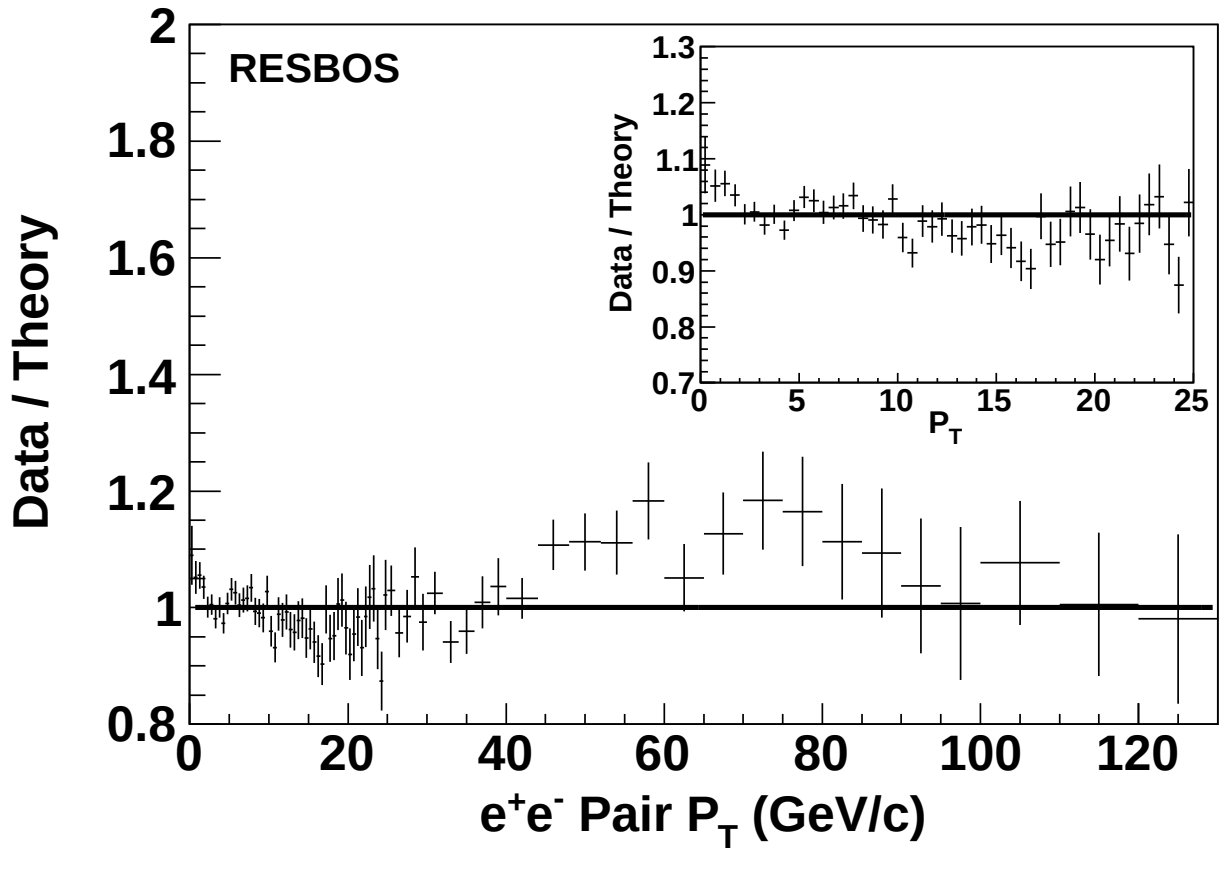

FIG. 8. The ratio of the measured cross section to the RESBOS prediction in the $P_{\mathrm{T}} < 130$ GeV/$c$ region. The RESBOS total cross section is normalized to the data. The inset is an expansion of the low $P_{\mathrm{T}}$ region.

TABLE I. The $\Delta\sigma/\Delta P_{\rm T}$ cross section versus $P_{\rm T}$. The first uncertainty is the statistical uncertainty. The second uncertainty is the efficiency measurement systematic uncertainty, which is 100% correlated across all bins. The 5.8% luminosity uncertainty applies to all bins but is not included.

| $P_{\rm T}$ bin GeV/$c$ | $\Delta\sigma/\Delta P_{\rm T}$ pb / GeV/$c$ | $P_{\rm T}$ bin GeV/$c$ | $\Delta\sigma/\Delta P_{\rm T}$ pb / GeV/$c$ |
|---|---|---|---|
| 0.0–0.5 | $(3.613 \pm 0.168 \pm 0.035) \times 10^0$ | 20.5–21.0 | $(2.923 \pm 0.143 \pm 0.030) \times 10^0$ |
| 0.5–1.0 | $(1.008 \pm 0.027 \pm 0.010) \times 10^1$ | 21.0–21.5 | $(2.877 \pm 0.144 \pm 0.030) \times 10^0$ |
| 1.0–1.5 | $(1.551 \pm 0.033 \pm 0.015) \times 10^1$ | 21.5–22.0 | $(2.603 \pm 0.134 \pm 0.027) \times 10^0$ |
| 1.5–2.0 | $(1.947 \pm 0.037 \pm 0.019) \times 10^1$ | 22.0–22.5 | $(2.624 \pm 0.139 \pm 0.027) \times 10^0$ |
| 2.0–2.5 | $(2.158 \pm 0.039 \pm 0.021) \times 10^1$ | 22.5–23.0 | $(2.590 \pm 0.140 \pm 0.026) \times 10^0$ |
| 2.5–3.0 | $(2.295 \pm 0.040 \pm 0.023) \times 10^1$ | 23.0–23.5 | $(2.516 \pm 0.139 \pm 0.026) \times 10^0$ |
| 3.0–3.5 | $(2.258 \pm 0.039 \pm 0.022) \times 10^1$ | 23.5–24.0 | $(2.200 \pm 0.124 \pm 0.022) \times 10^0$ |
| 3.5–4.0 | $(2.235 \pm 0.039 \pm 0.022) \times 10^1$ | 24.0–24.5 | $(1.948 \pm 0.113 \pm 0.020) \times 10^0$ |
| 4.0–4.5 | $(2.061 \pm 0.037 \pm 0.021) \times 10^1$ | 24.5–25.0 | $(2.179 \pm 0.129 \pm 0.022) \times 10^0$ |
| 4.5–5.0 | $(1.987 \pm 0.036 \pm 0.020) \times 10^1$ | 25.0–26.0 | $(2.032 \pm 0.085 \pm 0.021) \times 10^0$ |
| 5.0–5.5 | $(1.876 \pm 0.035 \pm 0.019) \times 10^1$ | 26.0–27.0 | $(1.736 \pm 0.076 \pm 0.018) \times 10^0$ |
| 5.5–6.0 | $(1.729 \pm 0.034 \pm 0.017) \times 10^1$ | 27.0–28.0 | $(1.633 \pm 0.075 \pm 0.016) \times 10^0$ |
| 6.0–6.5 | $(1.563 \pm 0.032 \pm 0.016) \times 10^1$ | 28.0–29.0 | $(1.616 \pm 0.077 \pm 0.016) \times 10^0$ |
| 6.5–7.0 | $(1.468 \pm 0.031 \pm 0.015) \times 10^1$ | 29.0–30.0 | $(1.381 \pm 0.069 \pm 0.014) \times 10^0$ |
| 7.0–7.5 | $(1.374 \pm 0.030 \pm 0.014) \times 10^1$ | 30.0–32.0 | $(1.284 \pm 0.045 \pm 0.013) \times 10^0$ |
| 7.5–8.0 | $(1.307 \pm 0.030 \pm 0.013) \times 10^1$ | 32.0–34.0 | $(1.005 \pm 0.039 \pm 0.010) \times 10^0$ |
| 8.0–8.5 | $(1.183 \pm 0.028 \pm 0.012) \times 10^1$ | 34.0–36.0 | $(8.769 \pm 0.361 \pm 0.088) \times 10^{-1}$ |
| 8.5–9.0 | $(1.112 \pm 0.027 \pm 0.011) \times 10^1$ | 36.0–38.0 | $(7.959 \pm 0.352 \pm 0.079) \times 10^{-1}$ |
| 9.0–9.5 | $(1.033 \pm 0.026 \pm 0.011) \times 10^1$ | 38.0–40.0 | $(7.068 \pm 0.336 \pm 0.070) \times 10^{-1}$ |
| 9.5–10.0 | $(1.024 \pm 0.027 \pm 0.011) \times 10^1$ | 40.0–44.0 | $(5.605 \pm 0.193 \pm 0.055) \times 10^{-1}$ |
| 10.0–10.5 | $(9.043 \pm 0.244 \pm 0.094) \times 10^0$ | 44.0–48.0 | $(4.600 \pm 0.179 \pm 0.044) \times 10^{-1}$ |
| 10.5–11.0 | $(8.295 \pm 0.231 \pm 0.084) \times 10^0$ | 48.0–52.0 | $(3.552 \pm 0.156 \pm 0.033) \times 10^{-1}$ |
| 11.0–11.5 | $(8.319 \pm 0.239 \pm 0.085) \times 10^0$ | 52.0–56.0 | $(2.760 \pm 0.136 \pm 0.025) \times 10^{-1}$ |
| 11.5–12.0 | $(7.780 \pm 0.229 \pm 0.079) \times 10^0$ | 56.0–60.0 | $(2.311 \pm 0.128 \pm 0.020) \times 10^{-1}$ |
| 12.0–12.5 | $(7.465 \pm 0.227 \pm 0.076) \times 10^0$ | 60.0–65.0 | $(1.618 \pm 0.089 \pm 0.014) \times 10^{-1}$ |
| 12.5–13.0 | $(6.839 \pm 0.215 \pm 0.069) \times 10^0$ | 65.0–70.0 | $(1.343 \pm 0.084 \pm 0.011) \times 10^{-1}$ |
| 13.0–13.5 | $(6.411 \pm 0.208 \pm 0.065) \times 10^0$ | 70.0–75.0 | $(1.094 \pm 0.078 \pm 0.009) \times 10^{-1}$ |
| 13.5–14.0 | $(6.220 \pm 0.208 \pm 0.064) \times 10^0$ | 75.0–80.0 | $(8.415 \pm 0.678 \pm 0.068) \times 10^{-2}$ |
| 14.0–14.5 | $(5.890 \pm 0.204 \pm 0.060) \times 10^0$ | 80.0–85.0 | $(6.347 \pm 0.565 \pm 0.049) \times 10^{-2}$ |
| 14.5–15.0 | $(5.363 \pm 0.190 \pm 0.055) \times 10^0$ | 85.0–90.0 | $(4.982 \pm 0.504 \pm 0.038) \times 10^{-2}$ |
| 15.0–15.5 | $(5.186 \pm 0.190 \pm 0.053) \times 10^0$ | 90.0–95.0 | $(3.786 \pm 0.422 \pm 0.028) \times 10^{-2}$ |
| 15.5–16.0 | $(4.792 \pm 0.181 \pm 0.049) \times 10^0$ | 95.0–100.0 | $(2.988 \pm 0.389 \pm 0.023) \times 10^{-2}$ |
| 16.0–16.5 | $(4.431 \pm 0.172 \pm 0.045) \times 10^0$ | 100.0–110.0 | $(2.298 \pm 0.227 \pm 0.016) \times 10^{-2}$ |
| 16.5–17.0 | $(4.149 \pm 0.165 \pm 0.042) \times 10^0$ | 110.0–120.0 | $(1.449 \pm 0.178 \pm 0.010) \times 10^{-2}$ |
| 17.0–17.5 | $(4.346 \pm 0.179 \pm 0.044) \times 10^0$ | 120.0–130.0 | $(9.369 \pm 1.389 \pm 0.064) \times 10^{-3}$ |
| 17.5–18.0 | $(3.931 \pm 0.166 \pm 0.040) \times 10^0$ | 130.0–140.0 | $(8.395 \pm 1.496 \pm 0.055) \times 10^{-3}$ |
| 18.0–18.5 | $(3.757 \pm 0.163 \pm 0.038) \times 10^0$ | 140.0–150.0 | $(5.304 \pm 1.174 \pm 0.034) \times 10^{-3}$ |
| 18.5–19.0 | $(3.753 \pm 0.167 \pm 0.038) \times 10^0$ | 150.0–175.0 | $(1.861 \pm 0.331 \pm 0.012) \times 10^{-3}$ |
| 19.0–19.5 | $(3.586 \pm 0.163 \pm 0.036) \times 10^0$ | 175.0–200.0 | $(5.283 \pm 1.478 \pm 0.031) \times 10^{-4}$ |
| 19.5–20.0 | $(3.303 \pm 0.154 \pm 0.034) \times 10^0$ | 200.0–250.0 | $(2.838 \pm 1.019 \pm 0.019) \times 10^{-4}$ |
| 20.0–20.5 | $(2.952 \pm 0.142 \pm 0.030) \times 10^0$ | 250.0–350.0 | $(1.489 \pm 1.162 \pm 0.009) \times 10^{-4}$ |

The $P_{\rm T}$ bins are numbered consecutively, 0–81 (lowest to highest $P_{\rm T}$), and the bin number is denoted by $n$. The bin-number expectation values of the eigenvectors are used for their assignment to $P_{\rm T}$ bins. The eigenvector with the lowest expectation value is assigned to $P_{\rm T}$ bin 0, the next lowest to $P_{\rm T}$ bin 1, and so on. For $P_{\rm T} < 25$ GeV/$c$, the rms width of the expectation value is about 4 bins, and above it, about 1 bin or less. In the 13–18 GeV/$c$ region, the rms width is the largest, 5–6 bins.

The $\chi^2$ is calculated for the eigenvector associated with the $P_{\rm T}$ bin $n$. For reference, the uncorrelated $\chi^2$ is also calculated for the bin. The cumulative $\chi^2$ from bin 0 to $n$ inclusive is denoted as $\chi^2(n)$. The number of degrees of freedom of $\chi^2(n)$ is $n$. A useful measure is $\chi^2(n) - n$: it is typically constant when the prediction is compatible with the data and increases over regions with discrepancies.

Figure 9 shows the $\chi^2(n) - n$ of the the RESBOS prediciton. For the correlated $\chi^2$, changes in $\chi^2(n) - n$ can only be associated with a $P_{\rm T}$ region because of smearing. In the $P_{\rm T} < 25$ GeV/$c$ region (bins 0–49), there are small differences but the data may allow further tuning of the RESBOS non-perturbative form factor that is important in this region.

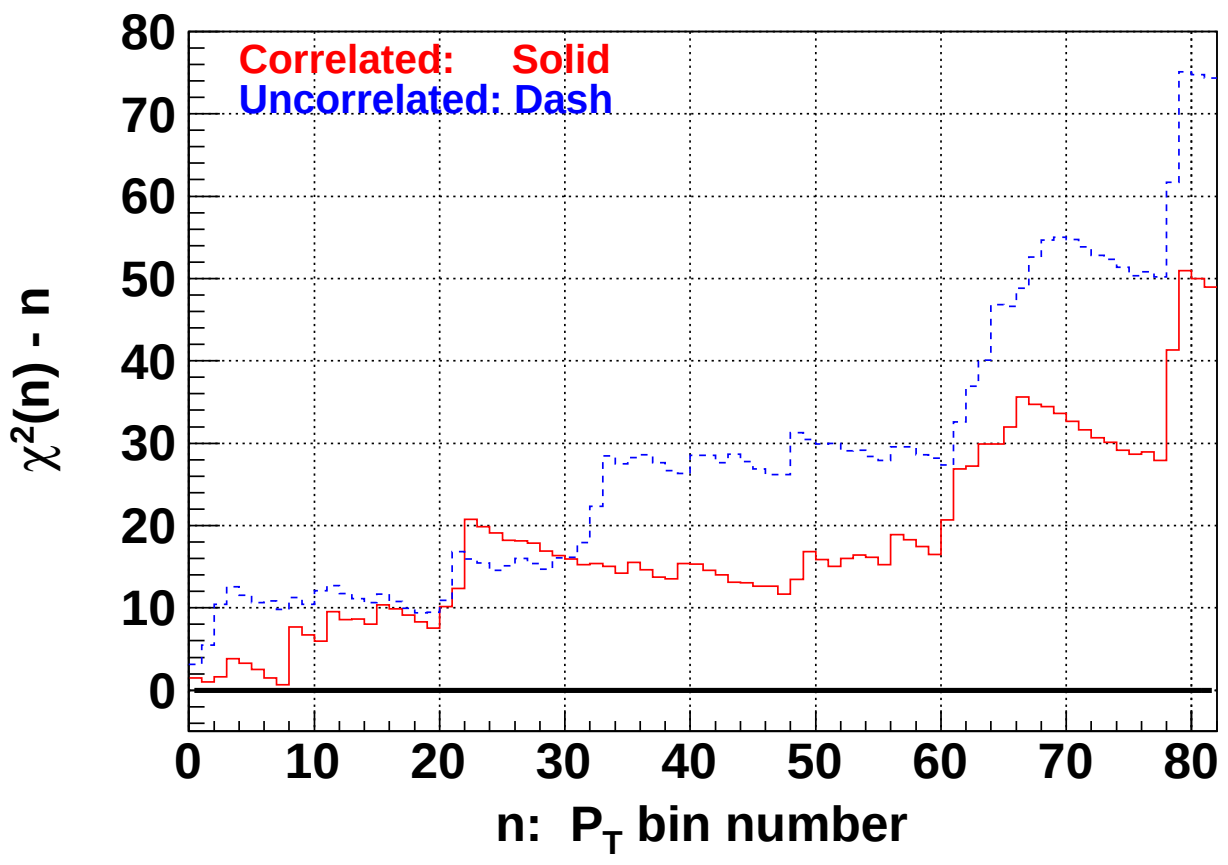

FIG. 9. $\chi^2(n)-n$ versus $P_{\rm T}$ bin number of the RESBOS prediction. The solid (red) histogram includes bin correlations and the dashed (blue) histogram does not. Bins 0–49 cover the 0–25 GeV/$c$ region, bin 60 is 40 GeV/$c$, bin 70 is 85 GeV/$c$, and bin 80 is 200 GeV/$c$.

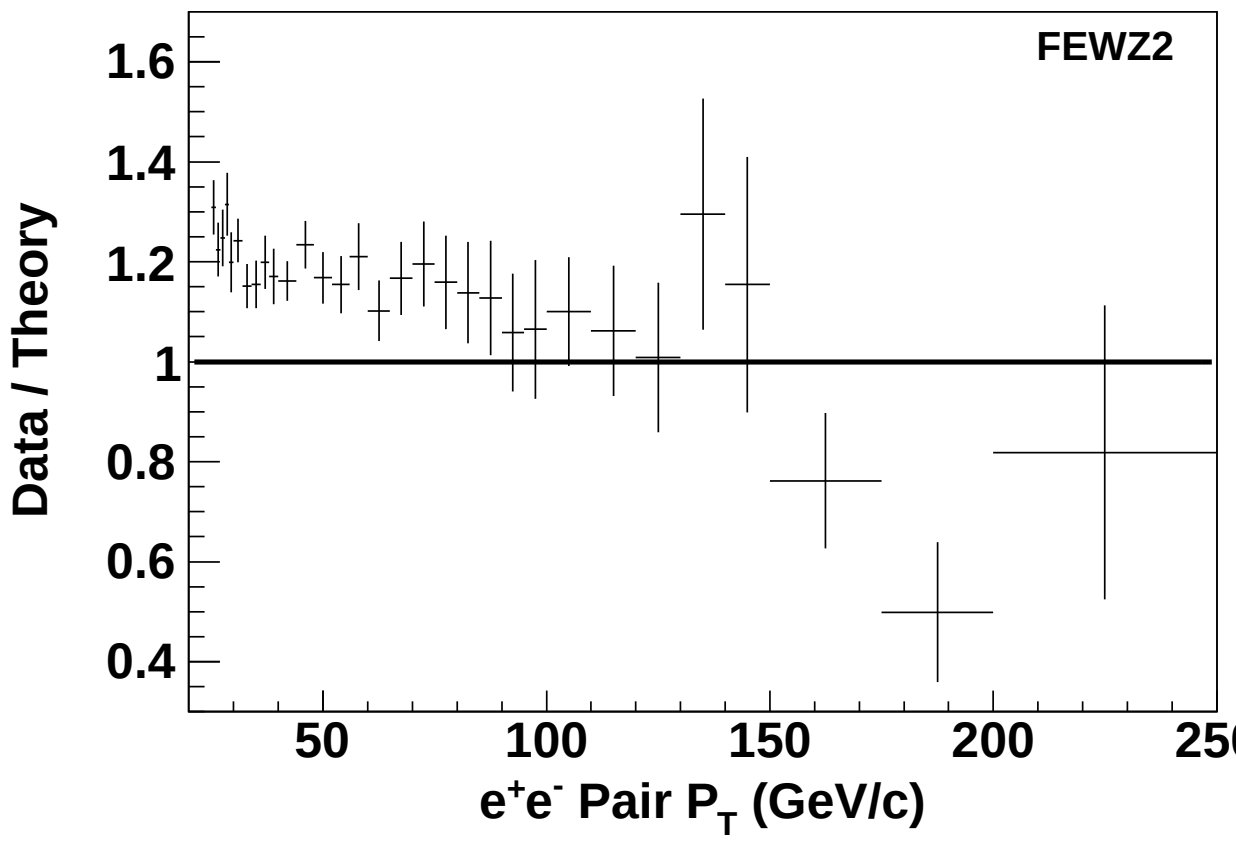

FIG. 10. The ratio of the measured cross section to the FEWZ2 prediction in the $25 < P_{\rm T} < 250$ GeV/$c$ region. The FEWZ2 cross section is not normalized to the data.

In the $44 < P_{\rm T} < 90$ GeV/$c$ region of Fig. 8 (bins 61–70 of Fig. 9), the RESBOS prediction is systematically lower than the data. This region is where the resummed calculation must be matched to the fixed-order perturbative calculation. This region is where the data can also contribute to the RESBOS resummation phenomenology of the Drell–Yan lepton pair $P_{\rm T}$ distribution at the Tevatron.

Figure 10 shows the ratio of the measured cross section to the FEWZ2 prediction. There is reasonable agreement with the data in the high $P_{\rm T}$ region where the RESBOS and FEWZ2 calculations are in agreement with each other. In $P_{\rm T}$ bins where the deviation of the FEWZ2 prediction from the measurement is significant, the difference provides a measure of the importance of higher order contributions above $\mathcal{O}(\alpha_s^2)$. The PDF uncertainties provided by FEWZ2 are at the 3% to 4% level. The uncertainties from variations of the QCD factorization and renormalization scales (from the $Z$-boson mass) in the $P_{\rm T}$ regions of 25–30, 100–110, and 200–250 GeV/$c$, are at the 7%, 5%, and 6% level, respectively. However, the accuracy of these scale uncertainties is unclear because of the two different scales (lepton-pair mass and transverse momentum) inherent in this QCD calculation.

## VII. SUMMARY

The transverse momentum cross section of $e^+e^-$ pairs in the $Z$-boson mass region of 66–116 GeV/$c^2$ produced in $p\bar{p}$ collisions is measured using 2.1 fb$^{-1}$ of Run II data collected by the Collider Detector at Fermilab. The measurement is data driven and corrected for the detector acceptance and smearing. The physics and detector models of the simulation used for the correction are tuned so that the simulation matches the data. The precision of the data and the measurement method require both the data and simulation to be well calibrated and understood. The measurement uncertainties are from a simulation-based model that quantifies the effects of event migration between measurement bins due to detector smearing.

Comparisons of this measurement with current quantum chromodynamic $\mathcal{O}(\alpha_s^2)$ perturbative and all-orders gluon resummation calculations show reasonable agreement. The data is of sufficient precision for further refinements in the phenomenology of the Drell–Yan lepton pair transverse momentum distribution.

## ACKNOWLEDGMENTS

We thank C.-P. Yuan for useful discussions and help on the CSS QCD resummation formalism and the RESBOS calculation. We thank F. Petriello for useful discussions and help on the FEWZ2 calculation. We thank the Fermilab staff and the technical staffs of the participating institutions for their vital contributions. This work was supported by the U.S. Department of Energy and National Science Foundation; the Italian Istituto Nazionale di Fisica Nucleare; the Ministry of Education, Culture, Sports, Science and Technology of Japan; the Natural Sciences and Engineering Research Council of Canada; the National Science Council of the Republic of China; the Swiss National Science Foundation; the A.P. Sloan Foundation; the Bundesministerium für Bildung und Forschung, Germany; the Korean World Class University Program, the National Research Foundation of Korea; the Science and Technology Facilities Council and the Royal Society, UK; the Russian Foundation for Basic Research; the Ministerio de Ciencia e Innovación, and Programa Consolider-Ingenio 2010, Spain; the Slo-vak R&D Agency; the Academy of Finland; and the Australian Research Council (ARC).

---